\documentclass[pre,twocolumn,superscriptaddress,showpacs,floatfix,amssymb,amsmath]{revtex4}

\usepackage[pdftex]{graphicx}% Include figure files
\usepackage{dcolumn}% Align table columns on decimal point
\usepackage{bm}% bold math
\usepackage{color}
\usepackage{amsmath}
\usepackage{amssymb}
\usepackage{mathrsfs}
\usepackage{epsfig}

\usepackage{array}
\usepackage{multirow}
\usepackage[justification=RaggedRight,singlelinecheck=false,font=small]{caption}
\usepackage{subfig}

\usepackage{bbm}

\newcommand{\la}[1]{\label{#1}}
\newcommand{\be}{\begin{eqnarray}}
\newcommand{\ee}{\end{eqnarray}}

\begin{document}
%

%\begin{center}{\large{REVISED}}\end{center}

\title{Thermal unfolding of myoglobin in the \\
Landau-Ginzburg-Wilson approach}

%\vskip 5.0cm
\author{Xubiao Peng}
\email{xubiaopeng@gmail.com}
\affiliation{Department of Physics and Astronomy, University of British Columbia, Vancouver, British Columbia V6T1Z4, Canada}
%\vskip 5.0cm
\author{Adam K. Sieradzan}
\email{adams86@wp.pl}
\affiliation{Faculty of Chemistry, University of Gdansk, Wita Stwosza 63, 80-308 Gda\'nsk, Poland}
\author{Antti J. Niemi}
\email{Antti.Niemi@physics.uu.se}
\affiliation{Department of Physics and Astronomy, Uppsala University,
P.O. Box 803, S-75108, Uppsala, Sweden}
\affiliation{
Laboratoire de Mathematiques et Physique Theorique
CNRS UMR 6083, F\'ed\'eration Denis Poisson, Universit\'e de Tours,
Parc de Grandmont, F37200, Tours, France}
\affiliation{Department of Physics, Beijing Institute of Technology, Haidian District, Beijing 100081, People's Republic of China}

\begin{abstract}
\noindent
The Landau-Ginzburg-Wilson paradigm is applied to
model the low-temperature crystallographic C$\alpha$ backbone structure of sperm 
whale myoglobin. The Glauber protocol is employed
to simulate its response  to an increase in ambient temperature. 
The myoglobin is found to unfold from its native state by a succession of $\alpha$-helical intermediates, 
fully in line with the observed  folding and unfolding patterns in denaturation experiments. 
In particular, a molten globule intermediate  is identified with experimentally correct attributes. 
A detailed, experimentally testable
contact  map is  constructed to characterise the specifics of the unfolding 
pathway,  including the formation of long range interactions. 
The results reveal how the unfolding process of a protein is driven by the interplay between, and 
a successive melting of,  its modular secondary structure components. 
\end{abstract}

\maketitle

\section{Introduction}

According to a paradigm by Anfinsen  \cite{anfinsen_1975}, under isothermal physiological conditions
the native structure of a protein relates to  the global minimum of Helmholtz free energy $F$ 
\begin{equation}
F=U-TS
\la{FU}
\end{equation}
Here $U$ is the internal energy, $S$ is the entropy and $T$ is the temperature.  
The Landau-Ginzburg-Wilson (LGW) approach 
\cite{landau_1937,widom_1965,kadanoff_1966,wilson_1971,fisher_1974,goldenfeld_1992,douglas_2009}
is a systematic method to approximate (\ref{FU}), in terms of
the symmetry properties of the underlying physical system. 
The approach was originally conceived to describe the static properties of phase 
transitions and critical phenomena. There, it has found numerous applications for example in
ordinary and quantum fluids, magnetic materials and superconductors. The approach 
reveals that independently of atomic level details, many {\it a priori} different material systems 
display identical universal behaviour,  when compared at sufficiently long spatial 
or temporal length scales.  Subsequently the LGW approach  has been 
expanded to describe time dependent critical phenomena. It has also 
been extended to model  {\it e.g.} pattern formation in non-equilibrium statistical systems and
chaotic behaviour in nonlinear dynamics \cite{goldenfeld_1992}.  Even aspects of fundamental string theory,
singularity theory and proof of existence of solutions to certain
nonlinear partial differential equations relate to the LGW approach \cite{douglas_2009}.

In the present article we develop and 
apply the Landau-Ginzburg-Wilson approach to model protein dynamics.
As an example we consider the way how 
myoglobin folds and unfolds when the ambient temperature increases.
 
Myoglobin is  the first protein to have its stable three-dimensional structure determined by x-ray 
crystallography \cite{kendrew_1958}. 
It  is one of the  most widely studied protein structures  \cite{alberts_2014}. 

The {\it protein folding problem} 
remains under an active scrutiny \cite{dill_2008,dill_2012,pettitt_2013}.  Many theoretical proposals 
have been presented, to explain how the folding of a protein
might proceed  \cite{lopez_1996,zhou_2004,bashford_1988,karplus_1994}. 
Recently, a soliton-based method which is built on the LGW approach 
has been presented, to describe both static folded proteins  \cite{chernodub_2010,molkenthin_2011} and 
aspects of protein dynamics \cite{krokhotin_2012a,krokhotin_2013a}.  
This method has been tested and validated computationally,  by comparing its predictions both with
a coarse-grained   \cite{krokhotin_2014,sieradzan_2014} and all-atom force-fields
\cite{ilieva_2015,dai_2015}. The simulations confirm that the folding 
of a simple protein proceeds by  a soliton formation, in a manner that can be accurately modeled using the 
Landau-Ginzburg-Wilson paradigm.

Here we combine the LGW approach 
with Glauber dynamics to study in detail, how myoglobin folds and unfolds. Glauber dynamics is
a Markov chain Monte Carlo method, that  is widely used to describe near-equilibrium 
relaxation dynamics of a statistical system towards equilibrium Gibbsian state \cite{glauber_1963,bortz_1975,berg_2004}. 

We note that in the case of a simple
spin system the Glauber protocol reduces to the Arrhenius relaxation law, and we also note that  the folding of simple
proteins appears to follow the Arrhenius law \cite{scalley_1997}. 

In the case of myoglobin, instead of temperature 
variations, most experiments have thus far 
utilised denaturants to study the unfolding and folding dynamics. However, in a 
computational approach it is more convenient to use the ambient temperature 
as the variable. 
Moreover, the experiments have mainly concentrated
on the heme-free apomyoglobin 
%Due to complications that are related to the binding of heme,
\cite{griko_1988,jennings_1993,shin_1993,eliezer_1995,eliezer_1996,jamin_1998,jamin_1999,uzawa_2004,nishimura_2006,uzawa_2008,meinhold_2011,nishimura_2011,xu_2012}. The heme containing
myoglobin has also been investigated 
\cite{hargrove_1996,culbertson_2010,ochiai_2010,moriyama_2010,ochiai_2011,uppal_2015} but due to apparent
complications with the binding of the heme, the studies have been limited to the  unfolding process.
In both cases the unfolding of the native state proceeds in stages 
with  several folding intermediates. In the case of apomyoglobin, the folding appears to proceed 
inversely to the unfolding.  
The dynamics is also very similar in both cases, 
except that in the apomyoglobin the F helix  \cite{alberts_2014} is initially disordered \cite{eliezer_1996} while in the heme containing myoglobin
the F helix is initially stable but  
the first to become disordered  when  temperature and/or denaturation increases
\cite{hargrove_1996,culbertson_2010,ochiai_2010,moriyama_2010,ochiai_2011,uppal_2015}. 
The unfolding of the F helix is followed by an intermediate molten globule, in both cases
\cite{griko_1988,jennings_1993,shin_1993,eliezer_1995,eliezer_1996,jamin_1998,jamin_1999,uzawa_2004,nishimura_2006,uzawa_2008,meinhold_2011,nishimura_2011,xu_2012,hargrove_1996,culbertson_2010,ochiai_2010,moriyama_2010,ochiai_2011,uppal_2015}.  
When denaturation and/or temperature increases further, 
the overall helicity of the molten globule rapidly decreases as the
helices B,C,D and E  start to unfold. 
Finally, the remaining helices A, G and H loose their stability 
and the structure becomes a random chain \cite{hargrove_1996,culbertson_2010,ochiai_2010,moriyama_2010,ochiai_2011,uppal_2015}. 

Here we show that the experimentally  observed unfolding pattern of myoglobin can be
%We model computationally the thermally driven unfolding process of myoglobin. in terms of standard
%non-equilibrium statistical physics.  In particular, we demonstrate that a 
accurately reproduced by a combination of Landau-Ginzburg-Wilson approach with  Glauber 
dynamics. Moreover, in line with the apomyoglobin experiments 
we show that the folding proceeds inversely to the unfolding. 
%is sufficient to fully reproduce  the experimentally observed unfolding pattern, 
%with a very high accuracy. 
In particular, we propose a contact map that describes the detailed order of helix formation, during both the 
unfolding and folding processes. Our predictions can be subjected to experimental tests, 
to reveal the extent of validity of the LGW approach.
%Recall that the Landau free energy is simply the universal  low momentum infrared  limit 
%approximation of the thermodynamical Gibbs free energy, which is consistent with all the symmetries
%that characterise the underlying physical system. 
%The Glauber protocol \cite{Berg_2004} is a widely used  to model the near-equilibrium relaxation 
%dynamics of a statistical system towards a Gibbsian equilibrium state.
%In the case of a simple spin system it reduces to the Arrhenius 
%relaxation law,  which is often followed by simple proteins  \cite{Scalley_1997}.
%Our results supports that the folding and unfolding process of a protein can be
%modeled, with a surprisingly high accuracy,  using universal concepts that have been originally developed and widely 
%applied to describe dynamical aspects of phase transitions and critical phenomena \cite{Niemi_2014,LesHouches}.
%In particular, the universality of our methodology proposes that it is applicable to a wide class of proteins.  
% The results are in good agreement with the experimental 
%observations \cite{all}.  In particular, the sequential unfolding pattern of $\alpha$-helices, 
%the structure of the molten globule folding intermediate, even the overall time scales that characterise 
%the major folding intermediates are similar with the experimental results. 
%

Our experimental reference conformation is
the Protein Data Bank (PDB) \cite{berman_2000} structure 1ABS \cite{schlichting_1994}
of wild type sperm whale heme containing myoglobin.  There are 154 
amino acids, indexed $i = 0... 153$ in the PDB file.  
The structure has been measured at a very low temperature 
$\sim$20 K with very small thermal B-factors;  in our approach
a high experimental accuracy is desirable since the model we develop can
describe the folded protein structure with sub-\AA ngstr\"om precision.

\section {Methods}

\subsection{Continuous curves}

For completeness, we start with a review of basic relations in curve geometry  \cite{hansonbook,kuipers}.
We consider a space curve $\mathbf x(s)\! : [0,L] \to   \mathbb R^3$ where  
$L$ is the total length of the curve  and 
$s \in [0,L]$ measures its proper length so that
\begin{equation}
|| \dot {\mathbf x} || = 1
\la{prole}
\end{equation}
The unit tangent vector is
\begin{equation}
\mathbf t \ = \   \dot {\mathbf x} \ \equiv \  \frac{ d \hskip 0.2mm \mathbf x (s)} {ds} 
\la{t}
\end{equation}
The unit binormal vector is
\begin{equation}
\mathbf b \ = \ \frac{ \dot {\mathbf x} \times \ddot {\mathbf x} } { || 
\dot {\mathbf x} \times \ddot {\mathbf x} || }
\la{b}
\end{equation}
and the unit normal vector is
\begin{equation}
\mathbf n = \mathbf b \times \mathbf t
\la{n}
\end{equation}
The orthonormal triplet ($\mathbf n, \mathbf b, \mathbf t$) defines a framing of the curve
that is subject to the Frenet equation \cite{hansonbook,kuipers}
\begin{equation}
\frac{d}{ds}\left(
\begin{matrix} 
{\bf n} \\
{\bf b} \\
{\bf t} \end{matrix} \right) =  \left( \begin{matrix}
0 & \tau & -\kappa  \\ -\tau & 0 & 0 \\  \kappa & 0 & 0 \end{matrix} \right) 
\left(
\begin{matrix} 
{\bf n} \\
{\bf b} \\
{\bf t} \end{matrix} \right) 
\la{contDS1}
\end{equation}
Here 
\begin{equation}
\kappa(s) \ = \ \frac{ || \dot {\mathbf x} \times \ddot {\mathbf x} || } { || \dot  {\mathbf x} ||^3 }
\la{kappa}
\end{equation}
is the curvature and
\begin{equation}
\tau(s) \ = \ \frac{ (\dot {\mathbf x} \times \ddot {\mathbf x}) \cdot {\dddot {\mathbf x} }} { || \dot {\mathbf x} \times \ddot {\mathbf x} ||^2 }
\la{tau}
\end{equation}
is the torsion. The fundamental theorem 
of space curves states that the shape of every sufficiently regular curve in three-dimensional space 
is completely determined by its curvature and torsion; the extrinsic and intrinsic geometries of a curve coincide. 
Thus, whenever $\kappa(s)$ and $\tau(s)$ are known, we can compute the  Frenet framing from
(\ref{contDS1}) and we can then proceed to compute the shape of the curve by integrating (\ref{t}).
Accordingly the curvature and the torsion are the (only) natural variables for constructing an energy function of the curve. 
In particular, the shape of a static curve should be computable, as a minimum of the pertinent 
energy function. 

Whenever  (\ref{prole}) is valid, the tangent vector is given by (\ref{t}). 
But when there is an inflection point {\it i.e.} a parameter value $s=s_0$ so that  
the curvature vanishes
\begin{equation}
\kappa(s_0) = || \ddot {\mathbf x} (s_0) || = 0
\la{inflection}
\end{equation}
the vectors $\mathbf n$ and $\mathbf b$ are not determined and the Frenet framing can not be introduced.
However, there are other ways to frame a curve, in a manner that extends continuously through an inflection point  
and more generally through straight segments of the curve. An example is the Bishop (parallel transport)
framing \cite{bishop_1974}.
 
\begin{figure}[h]
        \centering
                \includegraphics[width=0.35\textwidth]{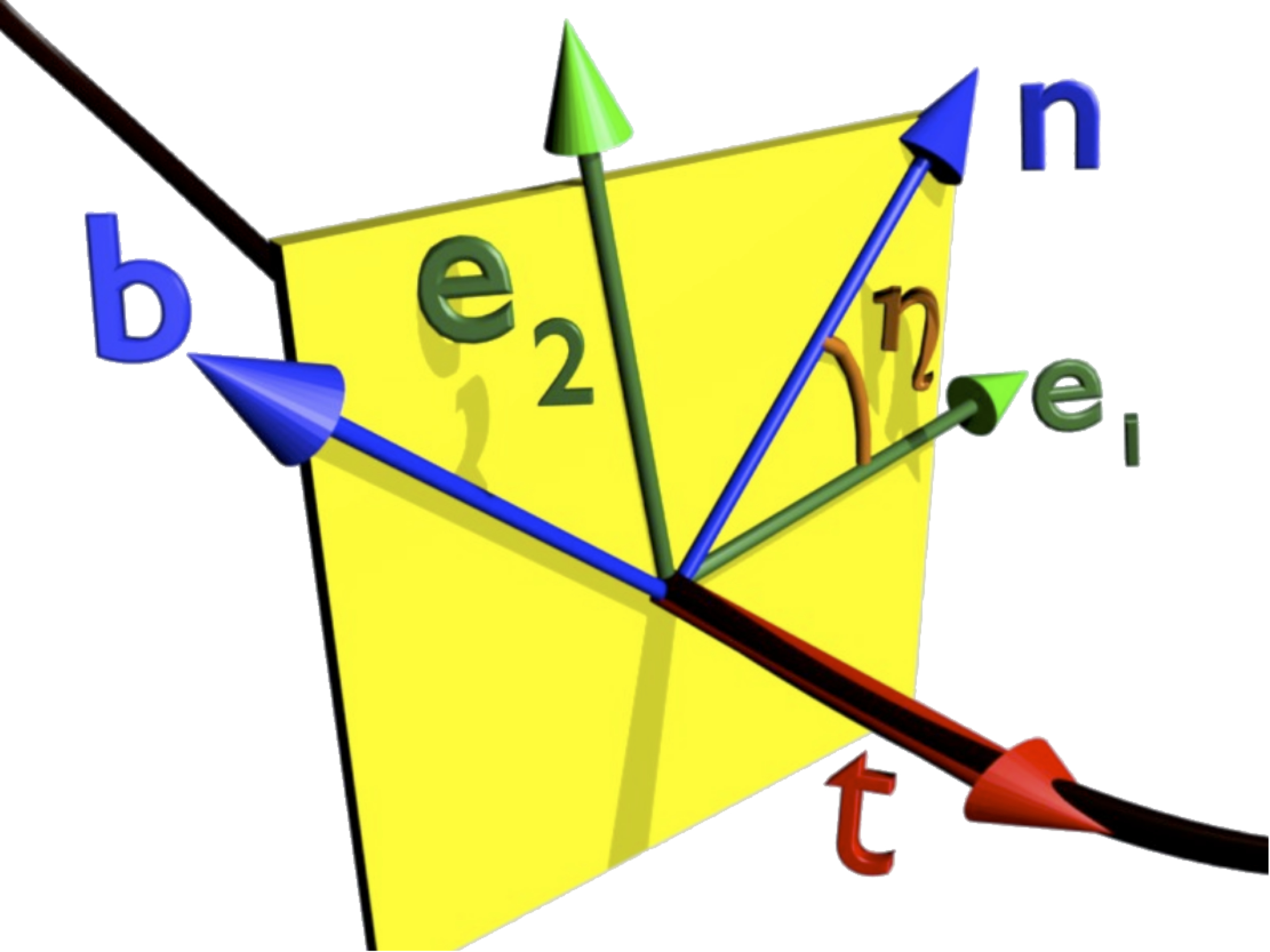}
        \caption{
      {{\it Color online:} The (blue) Frenet frame $(\mathbf n, \mathbf b)$ and a generic  (green) orthogonal frame $(\mathbf e_1 , \mathbf e_2)$ on the normal plane of $\mathbf t$,
      the tangent vector of the curve.}
                }
       \label{figure-1}
\end{figure}

The Landau-Ginzburg-Wilson paradigm states, that the energy function must be built so that it respects the
symmetries of the physical
system. For this we consider a generic orthonormal framing ($\mathbf e_1, \mathbf e_2, \mathbf t$). 
As shown in Figure \ref{figure-1} whenever the
curvature is non-vanishing it can be related to the Frenet framing by a local SO(2) rotation around the 
tangent vector $\mathbf t(s)$ 
%
%
%
%
%
%
%
%%%%%%%%%%%%%%%%%%%%%%%%%
%
%
%     Figure-1
%
%
%%%%%%%%%%%%%%%%%%%%%%%%%%
%
%
%
%
%
%
%
%
%
%%%%%%%%%%%%%%%%%%%%%%%%%
\begin{equation}
\left( \begin{matrix} {\bf n} \\ {\bf b} \end{matrix} \right) \ \to \ \left( \begin{matrix} {{\bf e}_1} \\ {\bf e}_2 \end{matrix} \right) \
= \ \left( \begin{matrix} \cos \eta(s) & - \sin \eta(s) \\ \sin \eta(s) & \cos \eta(s) \end{matrix}\right)
\left( \begin{matrix} {\bf n} \\ {\bf b} \end{matrix} \right)
\la{newframe}
\end{equation}
The ensuing generalisation of the Frenet equation is
\begin{equation}
\frac{d}{ds} \left( \begin{matrix} {\bf e}_1 \\ {\bf e }_2 \\ {\bf t} \end{matrix}
\right) =
\left( \begin{matrix} 0 & (\tau - \dot \eta) & - \kappa \cos \eta \\ 
- (\tau - \dot \eta)  & 0 & -\kappa \sin \eta \\
\kappa \cos \eta & \kappa \sin \eta  & 0 \end{matrix} \right)  
\left( \begin{matrix} {\bf e}_1  \\ {\bf e }_2 \\ {\bf t} \end{matrix}
\right) 
\la{contso2}
\end{equation} 
We deduce that the torsion transforms under frame rotations as follows,  
\begin{equation}
\tau \ \to \ \tau_r \ \equiv \ \tau - \dot \eta 
\la{sot}
\end{equation}
For the curvature, the effect of the frame rotation is summarised 
in terms of the complex valued quantity 
\begin{equation}
\kappa \ \to \ \kappa_\pm \ = \ \kappa e^{\pm i \eta} \ = \ \kappa_g \pm  i \kappa_n
\la{kot}
\end{equation}
The   (generalised)  Frenet equation can be represented  as follows,
\begin{equation}
(\frac{d}{ds} \pm  i \tau_r )  (\mathbf e_1 \pm i \mathbf e_2) \ \equiv \
(\frac{d}{ds} \pm  i \tau_r )  \mathbf e_\pm 
%\ = \ (\frac{d}{ds} \pm  i \tau_r )  \mathbf e_\pm  
= \ -  \kappa_\pm \bf t
 \la{e+-}
 \end{equation}
 \begin{equation}
\frac{d}{ds}  \mathbf t 
%\ = \ (\frac{d}{ds} \pm  i \tau_r )  \mathbf e_\pm  
= \ 2 ( \kappa_+ \mathbf  e_+ + \kappa_- \mathbf  e_-)
 \la{t-e}
 \end{equation}
The real part $\kappa_g$ of the complex curvature $\kappa_\pm$ 
is called the geodesic curvature, and the imaginary part $\kappa_n$ is called the normal curvature. These two
quantities refer to the extrinsic geometry of a surface that osculates the curve. The osculating surface is not uniquely determined, and 
different choices of $\eta$ correspond to different osculating surfaces.
The choice $\eta=0$ 
specifies the Frenet frame (Frenet gauge), and the choice
\[
\eta(s) =  \int_0^s \! \tau (s') ds' 
\]
specifies Bishop's frames \cite{hansonbook,kuipers,bishop_1974} that can be defined continuously and  unambiguously through an inflection point.

The invariance of the curve under frame rotations, {\it per se}, 
constitutes a symmetry that can be exploited to construct LGW energy functions.
We follow standard field theory  \cite{peskin_1995} and identify in
($\kappa_\pm, \tau_r$) a SO(2)$\sim$U(1) gauge 
multiplet (Abelian Higgs multiplet) \cite{niemi_2003}. The change (\ref{sot}) in $\tau_r $ 
is akin a SO(2)$\sim$U(1) gauge transformation of a one-dimensional gauge vector, 
while $\kappa_\pm$ transforms like  
a  complex scalar field. 

Finally, we observe that the complex valued Hashimoto variable \cite{hasimoto_1971} 
\begin{equation}
\xi(s) =  \kappa_+ (s) \exp\! \left(\! i\!\int_0^{s} \! \!\tau_r \,ds' \! \right) \equiv \kappa(s) \exp\! \left(\! i\! \int_0^{s} \! \! \tau\,ds'  \! \right)
\la{hash1}
\end{equation}
is gauge invariant {\it i.e.} independent of the choice of framing.

\subsection{Landau-Ginzburg-Wilson free energy }

The Landau-Ginzburg-Wilson approach  instructs us to exploit a  symmetry to construct 
an invariant energy function of a curve, in the limit of slow spatial variations.

We start with a generic Helmholtz free energy (\ref{FU}), in the limit 
of slow spatial variations; we follow \cite{coleman_1973}.
We assume a theory with a single scalar order parameter field $\varphi(x)$, {\it i.e.}
with no specific symmetry. 
The free energy (\ref{FU}) may be expanded in powers of the order parameter,
\begin{equation}
F =  \sum\limits_n \frac{1}{n !}\! \int\! d^D\! x_1 \cdot \cdot \cdot d^D\! x_n F^{(n)}(x_1 \cdot \cdot \cdot  x_n)\varphi(x_1) \cdot \cdot \cdot
\varphi(x_n)
\la{CW-1} 
\end{equation}
The coefficients $F^{(n)}$ are the $n$-point Green's functions, they are commonly 
evaluated perturbatively, in terms of Feynman diagrams  \cite{coleman_1973,peskin_1995}.

There is an alternative way to expand the free energy \cite{coleman_1973,peskin_1995},
in powers of derivatives (momentum) about the point where all external derivatives (momenta) vanish.
More specifically, we inspect the physical system over  
a distance scale $L$ such that the spatial variations of $\varphi(x)$ over this scale are small.
The  derivatives of $\varphi$ can then be employed as a small expansion parameters, and
\cite{coleman_1973,peskin_1995} 
\begin{equation}
F = \int\! d^D\! x \left [
V(\varphi) + \frac{1}{2} Z(\varphi) (\partial_\mu \varphi)^2 \cdot \cdot \cdot  \right]
\la{CW-2} 
\end{equation}
This is the
expansion in terms of slowly varying variables.  Note that
the coefficients $V(\varphi),  Z(\varphi), \dots$ are ordinary functions, not functionals. To the leading 
order $V(\varphi)$ coincides with
the classical potential in the Hamiltonian, generically \cite{coleman_1973,peskin_1995}
\[
 Z(\varphi) \ = \ 1 + a \varphi^2 + ...
 \]
 
We now specify to the case of a regular curve: The geometry of  a structureless curve 
can not depend on the way how it is framed, thus we propose to exploit
invariance under local frame rotations as the guiding symmetry;
the functional form of the ensuing Helmholtz free energy   (\ref{CW-2})
should remain intact under local frame rotations (\ref{newframe}).

In the case of a regular structureless curve, the 
shape  is completely determined by the generalised torsion (\ref{sot}) 
and curvature (\ref{kot}). Accordingly, these two local quantities 
constitute a complete set of order parameter variables, to 
specify the Helmholtz free energy  (\ref{CW-2}) of the curve.
Since the free energy should be independent of the way how the curve is framed,
it can only depend on gauge invariant {\it i.e.} frame rotation invariant
combinations of (\ref{sot}) and (\ref{kot}).
Thus, in the leading large distance (infrared) 
order the pertinent expansion (\ref{CW-2}) engages 
the Hamiltonian of the Abelian Higgs model \cite{niemi_2003,danielsson_2010}
\begin{equation}
F = \int \! ds \! \left [ \lambda\left(  |\kappa_+|^2 - m^2\right)^2 + |(\partial_s + i \tau_r)\kappa_+|^2 + \sigma \tau_r + \cdot \cdot \cdot \right]
\la{AHM-1}
\end{equation}
To the leading order, this is the most general non-local functional of ($\tau_r,\kappa_\pm$) 
which is manifestly invariant under the local frame rotation 
(\ref{newframe}).   

The last term in (\ref{AHM-1})  
is the {\it helicity}, it is a one dimensional version of the Chern-Simons term that breaks the chirality. 
The helicity is not  U(1) invariant, but its U(1)
transformation is a surface term.  Any surface term should 
become irrelevant in the thermodynamic limit. 

The variables ($\tau_r,\kappa_\pm$) can be eliminated in favour of the 
gauge invariant, geometric quantities (\ref{kappa}) and (\ref{tau}). This 
corresponds to the unitary gauge \cite{peskin_1995}:
We use (\ref{kot}) and 
\[
\tau \ = \ - \frac{i}{2\kappa^2} \left[ \kappa_- ( \partial_s + i \tau_r) \kappa_+ - c.c. \right]
\]
We substitute in (\ref{AHM-1}), and we obtain
\begin{equation}
F = \int ds \left [ (\partial_s \kappa)^2 + \kappa^2 \tau^2
+ \lambda( \kappa^2 - m^2)^2  + \sigma \tau \right]
\la{AHM-2}
\end{equation}

Specifically, the validity of the approximation (\ref{AHM-1}), (\ref{AHM-2}) 
assumes that  if $\kappa_0$ sets a  scale of curvature 
and when $L$ is a (large) distance scale of interest, then
\[
|\partial_s \kappa | <\! < \frac{\kappa_0}{L}  % \ \ \ \ \ \ \ \ \ (L >> \frac{1}{\kappa_0})
\]
Thus, over distance scales which are comparable to $L$ or larger, and subject to the frame rotation invariance,  
the Helmholtz free energy of a curve  is approximated by the LGW free energy, with leading
order expansion (\ref{AHM-2})
in derivatives of $\kappa(s)$.

\subsection{Integrable hierarchy}

Besides the frame rotation symmetry (\ref{newframe})-(\ref{kot}) there are other 
symmetry principles that may be utilised, as a guiding principle 
in the construction of a Landau-Ginzburg-Wilson
energy function. As an example, we consider the (infinite) symmetry which is 
associated with the concept of an integrable model  \cite{faddeev_2007,ablowitz_2004}: 

We start with the observation that the (manifestly frame independent)  Hasimoto variable (\ref{hash1}) 
converts the Hamiltonian 
of the Abelian 
Higgs Model into the Hamiltonian of the nonlinear Schr\"odinger (NLS) equation. Specifically, in terms of the variables 
(\ref{AHM-2})
\begin{equation}
\lambda \kappa^4 + \kappa^2 \tau^2 +  (\partial_s \kappa)^2  = \lambda(\bar \xi \xi)^2 + \partial_s \bar \xi \partial_s \xi 
\la{nlsh}
\end{equation}
which is the NLS hamiltonian \cite{faddeev_2007,ablowitz_2004,kevrekidis_2009}.
The NLS Hamiltonian is the paradigm integrable model. It 
admits an infinite number of conserved quantities, each associated with a  symmetry of (\ref{nlsh}).  
The last term in 
(\ref{AHM-2}), the helicity, is an example of a conserved quantity in the NLS model. The number density
\begin{equation}
\frac{1}{2} \bar\xi \xi = \frac{1}{2} \kappa^2 
\la{H2}
\end{equation}
is another example, and so is the momentum density
\begin{equation}
-\frac{i}{2} \bar \xi \partial_s \xi = \frac{1}{2} \kappa^2 \tau
\la{H1}
\end{equation}
Note that like helicity, momentum breaks chirality.

As such, (\ref{H2}) is the Hamiltonian of the  Worm Like Chain (Kratky-Porod) model \cite{kratky_1949}, 
widely used in modeling aspects of polymers.  In terms of the tangent vector,
\[
\frac{1}{2} \kappa^2  = \frac{1}{2}  |\partial_s \mathbf t|^2
\]
This is the Hamiltonian of the Heisenberg  $\sigma$-model \cite{faddeev_2007}. 

The LGW paradigm, in combination with the symmetry structure of the NLS model, 
proposes that the Helmholtz free energy  (\ref{FU}) can be systematically expanded in terms of the conserved
charges of the NLS hierarchy. In this way we arrive at the following (slight) generalisation of 
(\ref{AHM-2})
\[
F =  \int \! ds \! \left [ (\partial_s \kappa)^2  + \lambda( \kappa^2 - m^2)^2  + \right.
\]
\begin{equation}
 \left. + \frac{d}{2} \kappa^2 \tau^2  - 
b \kappa^2 \tau -  a \tau + \frac{c}{2} \tau^2 \right]
\la{finen}
\end{equation}
Here the last term is called the Proca mass in gauge theory, and we include it 
for completeness \cite{hu_2013,ioannidou_2014}.

The remaining conserved quantities of the NLS model involve higher order of derivatives of $\kappa(s)$. As such,
they are higher order corrections in the  expansion (\ref{CW-2}). We do not include them, in our infrared limit.

\subsection{Topological solitons}

Solitons are the paradigm structural
self-organisers in Nature  and the NLS equation is the paradigm 
equation that supports 
solitons \cite{faddeev_2007,ablowitz_2004,kevrekidis_2009};
depending on the sign of $\lambda$, the soliton 
is either dark ($\lambda >0$) or bright ($\lambda <0$).
Moreover, the torsion independent contribution to (\ref{finen}), (\ref{AHM-2})
\begin{equation}
 \int\limits_{-\infty}^\infty ds \, \left \{ \,  \kappa_s^2   + 
\lambda\, (\kappa^2 - m^2)^2 \, \right \}
\la{swave}
\end{equation}
supports  the double well  {\it topological} soliton \cite{manton_2004}:
When $m^2$ is positive and when $\kappa$ can take both positive and negative values, 
the equation of motion 
\[
\kappa_{ss} =  2 \lambda \kappa (\kappa^2 - m^2)
\]
is solved by 
\begin{equation}
\kappa(s) \ = \ m \, \tanh \left[ m \sqrt{\lambda} (s-s_0)\right]
\la{soliton}
\end{equation}
Note that this soliton engages an inflection point (\ref{inflection}); following  \cite{fadde,les-houches}
we use the convention that when a curve passes a simple 
inflection point, the curvature changes its sign.

The energy function (\ref{finen})  is quadratic in the torsion. Thus we can eliminate 
$\tau$ using its  equation of motion,
\begin{equation}
\tau [\kappa] \ = \ \frac{ a + b\kappa^2}{c+d\kappa^2} \ \equiv \ \frac{a}{c}  \, \frac{ 1 +  (b/a) \kappa^2}{1+(d/c)
\kappa^2}
\la{taueq}
\end{equation}
and we obtain the following equation of motion for curvature,
\begin{equation}
\kappa_{ss} =  V_\kappa [\kappa]
\la{monequ}
\end{equation}
where
\begin{equation}
V[\kappa] \ = \ - \left( \frac{ bc - ad}{d}\right) \, \frac{1}{c+d\kappa^2} \ - \ \left( \frac{b^2 + 8\lambda m^2}{2b} \right)
\, \kappa^2 + \lambda \, \kappa^4
\la{V}
\end{equation}
This shares the same large-$\kappa$ asymptotics, with the potential in (\ref{swave}).  
With properly chosen  parameters,   we expect that (\ref{monequ}), (\ref{V}) 
continue to support topological solitons.
But we do not know their explicit profile, in terms of elementary functions.

Once we have the soliton of (\ref{monequ}), we  evaluate $\tau(s)$ from  (\ref{taueq}).
We substitute the ensuing ($\kappa,\tau$) 
profiles in the Frenet equation (\ref{contDS1}) and solve for $\mathbf t(s)$. We then integrate
(\ref{t}) to obtain the curve $\mathbf x(s)$ that corresponds to the soliton. A generic soliton curve
looks like a helix-loop-helix motif, familiar from crystallographic protein structures. 
Note that depending on the parameter values, the torsion can fluctuate
substantially along a soliton profile even when curvature is slowly varying.

\subsection{Discrete Frenet equation}

Proteins can not be modeled by regular space curves. 
Proteins are like piecewise linear polygonal chains. In order to construct
the LGW energy function for a protein, we need to understand the structure and
symmetry of such a chain \cite{hu_2011}. 

Let $\mathbf r_i$ with $i=1,...,N$ be the vertices of a piecewise linear discrete chain;
in the case of a protein, the vertices correspond to  the C$\alpha$ atoms.
At each vertex we introduce the unit tangent vector 
\begin{equation}
\mathbf t_i = \frac{ {\bf r}_{i+1} - {\bf r}_i  }{ |  {\bf r}_{i+1} - {\bf r}_i | }
\la{dt}
\end{equation}
the unit binormal vector
\begin{equation}
\mathbf b_i = \frac{ {\mathbf t}_{i_1} - {\mathbf t}_i  }{  |  {\mathbf t}_{i_1} - {\mathbf t}_i  | }
\la{db}
\end{equation}
and the unit normal vector 
\begin{equation}
\mathbf n_i = \mathbf b_i \times \mathbf t_i
\la{dn}
\end{equation}
The orthonormal triplet ($\mathbf n_i, \mathbf b_i , \mathbf t_i$) constitutes a
discrete version of the Frenet  frames.

In lieu of the curvature and torsion, we have the  bond angles and torsion angles, defined as in
Figure \ref{fig_2}.
%%%%%%%%%%%%%%%%%%%%%%%%%%%%%%%%%%%%%%%%%%%%
%
%
%
%
%%%%%%%%%%%%%%%%%%%%%%%%%%%%%%%%%%%%%%%%%%%%%
%
%
%
%
%
%                           FIGURE 2
%
%
%
%
%
% 
%%%%%%%%%%%%%
%%%%%%%%%%%%%
%%%%%%%%%%%%%%
{
\footnotesize
\begin{figure}[h]         
 \centering      
  \resizebox{8 cm}{!}{\includegraphics[]{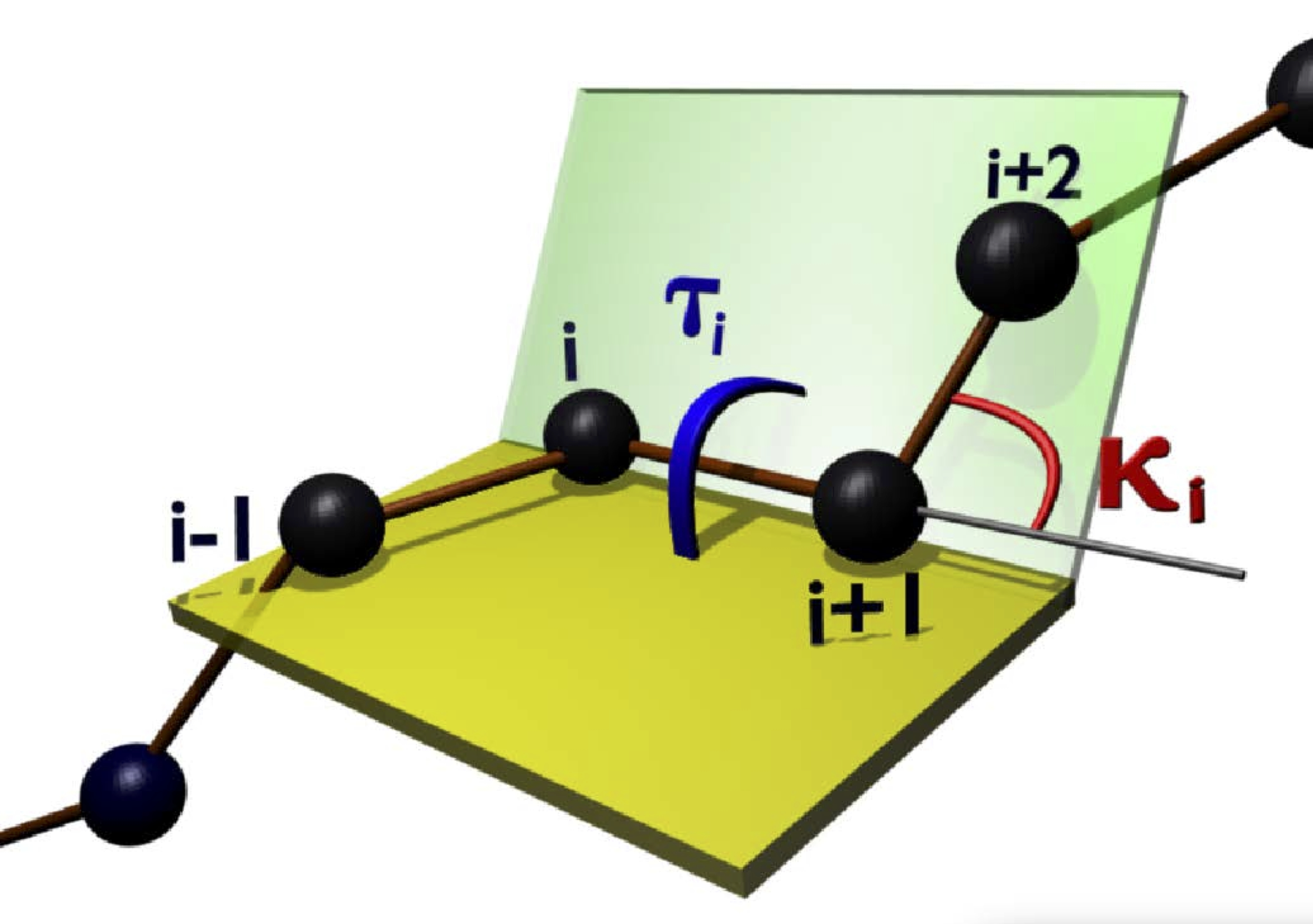}}
\caption {{\it Color online:} 
Definition of bond ($\kappa_i$) and torsion ($\tau_i$) angles, along  a piecewise linear discrete chain.   }   
\label{fig_2}    
\end{figure}
}
%%%%%%%%%%%%%
%%%%%%%%%%%%%
%%%%%%%%%%%%%%
%%%%%%%%%%%%%%%%%%%%%%%%%%%%%%%%%%%%%%%%%%%%
%
%
%
%
%%%%%%%%%%%%%%
%%%%%%%%%%%%%
%%%%%%%%%%%%%%

Once we know the Frenet frames at each vertex, we can compute the  angles. The bond angles are
\begin{equation}
\kappa_{i} \ \equiv \ \kappa_{i+1 , i} \ = \ \arccos \left( {\bf t}_{i+1} \cdot {\bf t}_i \right)
\la{bond}
\end{equation}
and the torsion angles are
\begin{equation}
\tau_{i} \ \equiv \ \tau_{i+1,i} \ = \ {\rm sign}\{ \mathbf b_{i_1} \times \mathbf b_i \cdot \mathbf t_i \}
\cdot \arccos\left(  {\bf b}_{i+1} \cdot {\bf b}_i \right) 
\la{tors}
\end{equation}
Conversely, when the values of the bond and torsion angles are all known, 
we can use the discrete Frenet equation
\begin{equation}
\left( \begin{matrix} {\bf n}_{i+1} \\  {\bf b }_{i+1} \\ {\bf t}_{i+1} \end{matrix} \right)
= 
\left( \begin{matrix} \cos\kappa \cos \tau & \cos\kappa \sin\tau & -\sin\kappa \\
-\sin\tau & \cos\tau & 0 \\
\sin\kappa \cos\tau & \sin\kappa \sin\tau & \cos\kappa \end{matrix}\right)_{\hskip -0.1cm i+1 , i}
\left( \begin{matrix} {\bf n}_{i} \\  {\bf b }_{i} \\ {\bf t}_{i} \end{matrix} \right) 
\la{DFE2}
\end{equation}
to  compute  the frame at vertex $i+i$ 
from the frame at vertex $i$. Once  all the frames have been constructed,  
the entire chain is given by
\begin{equation}
\mathbf r_k = \sum_{i=0}^{k_1} |\mathbf r_{i+1} - \mathbf r_i | \cdot \mathbf t_i
\la{dffe}
\end{equation}
Without any loss of generality we may choose $\mathbf r_0 = 0$, choose $\mathbf t_0$ to 
point into the direction of the positive $z$-axis, and let $\mathbf t_1$ lie on the $y$-$z$ plane.

As in the case of a continuum curve, a discrete chain remains intact under frame
rotations of  the ($\mathbf n_i, \mathbf b_i$) zweibein around $\mathbf t_i$.
This local SO(2)  rotation acts on the frames as follows
\begin{equation}
 \left( \begin{matrix}
{\bf n} \\ {\bf b} \\ {\bf t} \end{matrix} \right)_{\!i} \!
\rightarrow  \!  e^{\Delta_i T^3} \left( \begin{matrix}
{\bf n} \\ {\bf b} \\ {\bf t} \end{matrix} \right)_{\! i} =   \left( \begin{matrix}
\cos \Delta_i & \sin \Delta_i & 0 \\
- \sin \Delta_i & \cos \Delta_i & 0 \\ 
0 & 0 & 1  \end{matrix} \right) \left( \begin{matrix}
{\bf n} \\ {\bf b} \\ {\bf t} \end{matrix} \right)_{\! i}
\la{discso2}
\end{equation}
where $T^3$ is one of the SO(3) Lie algebra generators, 
\[
(T^a)_{bc} = \epsilon^{abc}
\]
In terms of the bond and torsion angles the rotation amounts to
\begin{equation}
\kappa_{i}  \ T^2  \ \to \  e^{\Delta_{i} T^3} ( \kappa_{i} T^2 )\,  e^{-\Delta_{i} T^3}
\la{sokd}
\end{equation}
\begin{equation}
\tau_{i}  \ \to \ \tau_{i} + \Delta_{i_1} - \Delta_{i}
\la{sotd}
\end{equation}
which is a direct generalisation of (\ref{sot}), (\ref{kot}); following standard field theory \cite{peskin_1995}
the transformation of bond angles is like an adjoint SO(2)$\in$SO(3) 
gauge rotation of a Higgs triplet around the Cartan
generator $T^3$, when the Higgs triplet is  in the (unitary gauge)
direction of $T^2$. The transformation of torsion angle coincides with that of the 
SO(2) lattice gauge field. 

{\it A priori}, the fundamental range of the bond angle is  $\kappa_i \in [0,\pi]$ while for the 
torsion angle the range is $\tau_i \in [-\pi, \pi)$. Thus we may
identify ($\kappa_i, \tau_i$) as the canonical 
latitude and longitude angles of a two-sphere $\mathbb S^2$. 
However, to account for the presence of putative inflection, it is useful to extend the range
of $\kappa_i$ into negative values $ \kappa_i \in [-\pi,\pi]$ $mod(2\pi)$.
We compensate for this two-fold covering of $\mathbb S^2$ 
by a $\mathbb Z_2$ symmetry:
\begin{equation}
\begin{matrix}
\ \ \ \ \ \ \ \ \ \kappa_{k} & \to  &  - \ \kappa_{k} \ \ \ \hskip 1.0cm  {\rm for \ \ all} \ \  k \geq i \\
\ \ \ \ \ \ \ \ \ \tau_{i }  & \to &  \hskip -2.5cm \tau_{i} - \pi 
\end{matrix}
\la{dsgau}
\end{equation}
This is a special case of (\ref{sokd}), (\ref{sotd}), with
\[
\begin{matrix} 
\Delta_{k} = \pi \hskip 1.0cm {\rm for} \ \ k \geq i+1 \\
\Delta_{k} = 0 \hskip 1.0cm {\rm for} \ \ k <  i+1 
\end{matrix}
\]

\subsection{The C$\alpha$ trace reconstruction}
%\label{sect12}

The discrete Frenet equation (\ref{DFE2}), (\ref{dffe}) discloses, that a chain 
can be constructed from the knowledge of bond and torsion angles and the distances 
between the vertices. In the case of crystallographic protein structures, the vertices 
coincide with the positions of the C$\alpha$ atoms. As shown in Figure \ref{fig-3}, 
in PDB the virtual C$\alpha$-C$\alpha$ 
bond lengths are very close to their average value 
\begin{equation}
 |\mathbf r_{i+1} - \mathbf r_i |  \ \sim \ 3.8 \ \ {\rm \AA}
\la{dist}
\end{equation}
%%%%%%%%%%%%%%%%%%%%%%%%%%%%%%%%%%%%%%%%%%%%
%
%
%
%%%\onecolumngrid
%%%
%%%\begin{center}
%%%\begin{figure}[h]
%%%\includegraphics[scale=1.0]{myfig.pdf}
%%%\caption{mycaption}
%%%\end{figure}
%%%\end{center}
%%%\twocolumngrid
%
%
%                           FIGURE 3
%
%
%
%
%
% 
%%%%%%%%%%%%%
%%%%%%%%%%%%%
%%%%%%%%%%%%%%
%
{
\footnotesize
\begin{figure}[h]         
\centering            
  \resizebox{8cm}{!}{\includegraphics[]{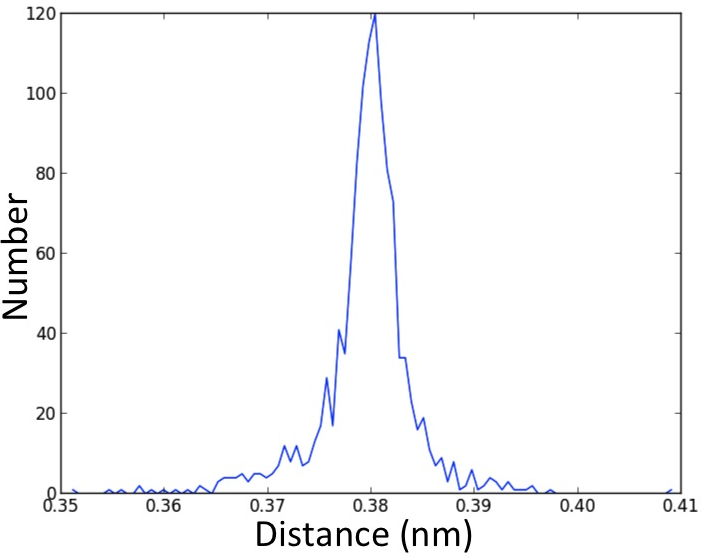}}
\caption {{\it Color online:} Distribution of bond length in crystallographic PDB structures; the data set in \cite{hinsen_2013} has been used. }   
\label{fig-3}    
\end{figure}
}
%
%%%%%%%%%%%%%
%%%%%%%%%%%%%
%%%%%%%%%%%%%%
%%%%%%%%%%%%%
%%%%%%%%%%%%%
%%%%%%%%%%%%%%
Moreover, according to  \cite{hinsen_2013}
the C$\alpha$ backbones of PDB structures
can be reliably reconstructed using a combination of the actual bond and torsion angles
(\ref{bond}), (\ref{tors}) and the {\it average} value (\ref{dist}). Thus the bond and torsion angles constitute
a complete set of structural order parameters, in the case of crystallographic proteins. The LGW paradigm
proposes that the leading order approximation to the 
Helmholtz free energy is a function of the bond and torsion angles only.

{\it Note:} The Ramachandran angles, together with the average value  (\ref{dist}), do {\it not} constitute
a complete set of structural order parameters \cite{hinsen_2013}.

\subsection{LGW Hamiltonian for proteins} 

Proteins are commonly modeled using an all-atom force field, or  a
coarse-grained approximation thereof \cite{dill_2008,dill_2012,pettitt_2013}.
The discretised Newton's equation is solved iteratively,  
in what {\it de facto} amounts to a perturbative 
expansion around a (randomly) chosen initial configuration:
The expansion parameter relates to the ratio of the iterative
time step length to the time scale of a characteristic atomic oscillation.  
In an all-atom approach the latter
pertains to the frequency of a heavy atom
covalent bond oscillation, which makes simulations into an 
extreme computational challenge.
The equation
(\ref{CW-1}) exemplifies a perturbative approach.

Here we follow the Landau-Ginzburg-Wilson paradigm to 
develop a complementary  approach to model proteins and their dynamics. 
Conceptually,  our approach is 
like  the expansion (\ref{CW-2}) and we need to identify the slowly varying variable:
In the Figure \ref{fig-4} we show the distribution of C$\alpha$ backbone bond and 
torsion angles in crystallographic PDB protein structures,
on the stereographically projected two-sphere ($\kappa,\tau$). The torsion angles 
%%%%%%%%%%%%%%%%%%%%%%%%%%%%%%%%%%%%%%%%%%%%
%
%
%
%
%
%                           FIGURE 4
%
%
%
%
%
% 
%%%%%%%%%%%%%
%%%%%%%%%%%%%
%%%%%%%%%%%%%%
{
\footnotesize
\begin{figure}[h]         
\centering            
  \resizebox{8 cm}{!}{\includegraphics[]{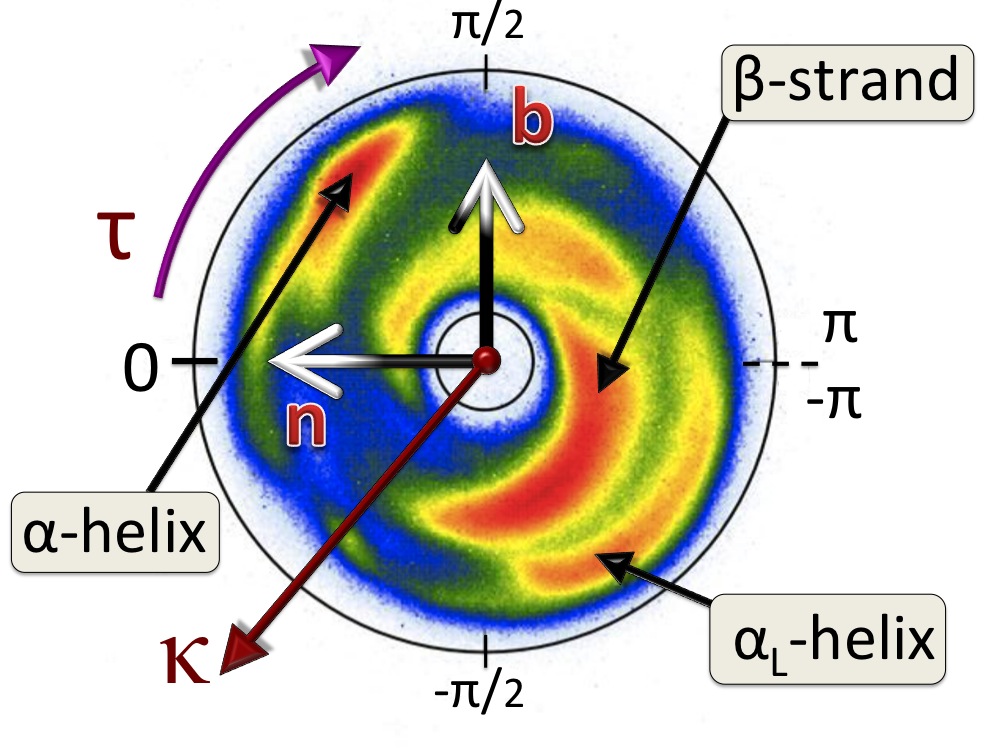}}
\caption {{\it Color online:} Distribution of bond and torsion angles in crystallographic protein structures, on a stereographically projected
two-sphere with $\kappa$ the latitude and $\tau$ the longitude.  Red indicates a large number of entries, blue a small number of entries, and white corresponds to no entries. All PDB structures that have been measured with resolution 2.0 \AA~ or better, have 
been used. The major secondary structure regimes are identified.
 The inner boundary of the annulus has a radius $\kappa \approx 1$ (rad) and the outer boundary has a radius $\kappa \approx 1.6$ (rad).}   
\label{fig-4}    
\end{figure}
}
%%%%%%%%%%%%%
%%%%%%%%%%%%%
%%%%%%%%%%%%%%
%%%%%%%%%%%%%
%%%%%%%%%%%%%
%%%%%%%%%%%%%%
are known to be flexible in proteins; as shown in Figure \ref{fig-4}, their values 
are distributed over the entire range $\tau \in (-\pi, \pi]$. However, the observed range of variation
$\Delta \kappa_{max}$ in the values of the bond angles is quite constrained. Instead of extending over the
entire plane, the angles are largely limited to the annulus between
$\kappa \approx 1$ and $\kappa \approx \pi/2$ (radians), shown in Figure \ref{fig-4}.  
On the original two-sphere, the geometrically
allowed range of variations of the bond angle is $\kappa_{tot} \in [0,\pi]$. Thus we may
putatively adopt the ratio
\begin{equation}
\left | \frac{ \kappa_{i+1} - \kappa_i}{\kappa_{tot}}\right | 
\leq \frac{\Delta \kappa_{max}}{\kappa_{tot}} \ \approx \ \frac{0.6}{\pi} \ \sim 0.2
\la{expar}
\end{equation}
as a slowly varying expansion parameter, in the case of crystallographic protein structures.  
The Landau-Ginzburg-Wilson paradigm then proposes that {\it  if} we adopt the NLS hierarchy 
as the symmetry principle to guide the construction of the LGW energy function,  in the case of a protein
backbone we should  adopt a discretised version of (\ref{finen})
as the leading order LGW approximation of the Helmholtz free energy: 
\[
H = - \sum\limits_{i=1}^{N_1}  2\, \kappa_{i+1} \kappa_{i}  + 
\sum\limits_{i=1}^N
\biggl\{  2 \kappa_{i}^2 + \lambda\, (\kappa_{i}^2 - m^2)^2  \biggr.
\]
\begin{equation}
\biggl. + \frac{d}{2} \, \kappa_{i}^2 \tau_{i}^2   
- b \, \kappa^2_i \tau_{i}   - a \,  \tau_{i} +  \frac{c}{2}  \tau^2_{i} 
\biggr\} \ + \dots
\la{E1old}
\end{equation}
The approximation (\ref{E1old}) should be a valid one, as long as the expansion parameter (\ref{expar}) remains small,
{\it i.e.} there are 
no abrupt but only slowly
changing bends along 
the backbone. In particular, long range interactions 
are accounted for, as long as they do not cause any sharp  localised 
buckling of the backbone.

In  (\ref{E1old}) $\lambda$, $a$, $b$, $c$, $d$ and $m$ 
depend on the atomic level physical properties and the chemical 
microstructure of the protein and its environment. In principle, these parameters can 
be computed from this knowledge.  In practice, we train the energy function to model a given protein.
 
 \subsection{Topological soliton and protein geometry}
 
The free energy (\ref{E1old}) is a {\it naive} discretisation of the NLS hierarchy 
free energy (\ref{finen}). It is a deformation of the energy function of the 
integrable discrete 
nonlinear Schr\"odinger  equation (DNLS)  \cite{faddeev_2007,ablowitz_2004,kevrekidis_2009}. 
The conventional DNLS equation is known to
support solitons. Thus we expect that (\ref{E1old}) supports soliton solutions as well:

We follow (\ref{taueq}) to eliminate the torsion angle, 
\begin{equation}
\tau_i [\kappa] \ = \ \frac{ a + b\kappa_i^2}{c+d\kappa_i^2} \ =  \ a  \frac{ 1 + b\kappa_i^2}{c+d\kappa_i^2}
\la{taueq2}
\end{equation}
For bond angles we then have
\begin{equation}
\kappa_{i+1} = 2\kappa_i - \kappa_{i_1} + \frac{ d V[\kappa]}{d\kappa_i^2} \kappa_i  \ \ \ \ \ (i=1,...,N)
\la{dnlse}
\end{equation}
We set $\kappa_0 = \kappa_{N+1}=0$, and $V[\kappa]$ is given by (\ref{V}). To solve this numerically,  
we use  the 
iterative equation \cite{molkenthin_2011}
\begin{equation}
\kappa_i^{(n+1)} \! =  \kappa_i^{(n)} \! - \epsilon \left\{  \kappa_i^{(n)} V'[\kappa_i^{(n)}]  
- (\kappa^{(n)}_{i+1} - 2\kappa^{(n)}_i + \kappa^{(n)}_{i_1})\right\}
\la{ite}
\end{equation}
where $\{\kappa_i^{(n)}\}_{i\in N}$ is the $n^{th}$ iteration of an initial configuration  
$\{\kappa_i^{(0)}\}_{i\in N}$ and $\epsilon$ is some sufficiently small but otherwise arbitrary 
numerical constant. We choose $\epsilon = 0.01$, in our simulations. The fixed point
of (\ref{ite}) is independent of the value of $\epsilon$,  and clearly a solution of (\ref{dnlse}).

Once the fixed point is found, the corresponding 
torsion angles are obtained from (\ref{taueq2}). The frames are then constructed from (\ref{DFE2}), and
the entire chain is constructed 
using  (\ref{dffe}).

We do not know of an analytical expression of the soliton solution 
to the equation (\ref{dnlse}). 
But an {\it excellent} approximative solution can be obtained  by discretizing the topological 
soliton (\ref{soliton})  \cite{chernodub_2010,fadde,les-houches,hu_2011,krokhotin_2011}:
\begin{equation}
\kappa_i \ \approx \   \frac{ 
m_{1}  \cdot e^{ c_{1} ( i-s) } - m_{2} \cdot e^{ - c_{2} ( i-s)}  }
{
e^{ c_{1} ( i-s) } + e^{ - c_{2} ( i-s)} }
\label{An1}
\end{equation}
Here ($c_1, c_2, ,m_{1},m_{2},s$) are parameters. The $m_{1}$ and $m_{2}$ 
specify the asymptotic $\kappa_i$-values of the soliton. Thus, 
these parameters are entirely determined by the character 
of the regular, constant bond and torsion angle structures that are adjacent to the 
soliton. In particular, these parameters are not specific  to the soliton {\it per se}, but to the adjoining  
regular structures.
The parameter $s$ defines the location of the soliton along the 
string.  This leaves us with only two loop specific parameter, the $c_{1}$ and $c_{2}$. 
These parameters quantify the length of the bond angle profile that describes the soliton. 

For the torsion angle, (\ref{taueq2}) involves one parameter ($a$) that we have factored out as the overall relative scale
between the bond angle and torsion angle contributions to the energy; this parameter determines the relative
flexibility of the torsion angles, with respect to the bond angles. Then, there are three additional
parameters ($b/a, c/a, d/a$)  in the remainder $\hat \tau[\kappa]$. Two of these are again determined by the character 
of the regular structures that are adjacent to the soliton. 
As such, these parameters are not specific  to the soliton. The remaining single parameter
specifies the size of the regime where the torsion angle fluctuates.

On the regions adjacent to a soliton, 
we have constant values of $(\kappa_i, \tau_i)$.  In the case of a protein, these are the regions 
that correspond to the standard regular secondary structures. For example, the 
standard right-handed $\alpha$-helix is obtained by setting
\begin{equation}
\alpha-{\rm helix:} \ \ \ \ \left\{ \begin{matrix} \kappa \approx \frac{\pi}{2}  \\ \tau \approx 1\end{matrix} \right.
\label{bc1}
\end{equation}
and for the standard $\beta$-strand 
\begin{equation}
\beta-{\rm strand:} \ \ \ \ \left\{ \begin{matrix} \kappa \approx 1 \\ \tau \approx \pi \end{matrix}  \right.
\label{bc2}
\end{equation}
All the other standard  regular secondary structures of proteins such as 3/10 helices, 
left-handed helices {\it etc.} are similarly modeled by definite constant values of $\kappa_i$ and $\tau_i$.
Protein loops correspond to solitons, the regions where the values of ($\kappa_i, \tau_i$) 
are variable. 

\vskip 0.3cm

\noindent
{\it The presence of solitons {\it significantly} reduces the number of parameters in (\ref{E1old}). The
number of parameters is far smaller than the number of amino acids, along the protein backbone. }

\subsection{Proteins out of thermal equilibrium}
\label{sect17}

When a protein folds towards its native state, it is out of thermal equilibrium. Several studies propose, that
in the case of a small protein which is not too far away from thermal equilibrium, 
the folding  takes place in a manner which is consistent with
Arrhenius' law \cite{scalley_1997}. This law states that the reaction
rate depends exponentially on the ratio of activation energy $E_A$ and physical temperature factor,
\[
r  \ \propto  \ \exp\{ - \frac{E_A}{k_B \theta}\}
\]
with $k_B$ the Boltzmann 
constant and $\theta$ the temperature measured in  Kelvin. 

On the other hand, in the case of  a simple spin chain, Glauber dynamics
\cite{glauber_1963,bortz_1975,berg_2004} is known to
describe the approach to thermal equilibrium, in a manner which resembles Arrhenius's law. 
Glauber dynamics evaluates the transition probability from a conformational state $a$ to another 
conformational state $b$  as follows:
\[
\mathcal P(a \to b) \ = \ \frac{1}{ 1 + e^{\Delta F_{ba} /T} }  
\]
Here $\Delta F_{ba} = F_b - F_a$ is the activation energy and  we compute it from  (\ref{E1old}). The parameter
$T$ is the Monte Carlo temperature factor. Note that in general
the Monte Carlo temperature factor $T$ does not coincide with the 
physical temperature factor $k_B \theta$.  Instead  
 we expect \cite{krokhotin_2013a} that $T$ relates to $k_B \theta$ approximatively as follows,
\begin{equation}
T \ \sim \ \alpha \, k_B \theta \, e^{\beta k_B \theta}
\la{Ttheta}
\end{equation}
where $\alpha$ and $\beta$ are protein specific factors.

\subsection{Simulation details}

In all our simulations, at each MC step we perturb either one of the bond angles or torsion angles according to the
following prescription:
\[
\begin{matrix} \kappa_i & \to & \kappa_i + 0.015 r \\ 
\tau_i & \to & \tau_i + 1.5 r ~~~
\end{matrix}
\]
where $r$ is a random number with Gau\ss ian distribution with expectation value 0
and variation 1. The different scales on $\kappa_i$ and $\tau_i$ reflect the different stiffness 
between the bond and torsion angles, in real proteins.
We have tested various other values of $r$,
to confirm that our results do not essentially depend on the choice of $r$.

In addition, in the case of a protein we need to account for
steric constraints: In PDB, for two C$\alpha$ atoms which are {\it not} nearest neighbours
along the backbone, we have 
\[
| \mathbf r_i - \mathbf r_k | > 3.8 \ {\mathrm \AA} \ \ \ {\rm for} \ \ \ |i-k| \geq 2
\]
We introduce this condition as a requirement, to accept a given
Monte Carlo step during simulation.

\section{Results}

We have performed extensive numerical simulations to analyse 
the way how myoglobin {\it unfolds} when ambient
temperature increases. The motivation to consider in detail 
the {\it unfolding} process in the case of a myoglobin, relates to the experimental 
issues due to the binding of heme: It is very difficult to control the process of heme binding
in a folding experiment, thus the myoglobin experiments 
\cite{hargrove_1996,culbertson_2010,ochiai_2010,moriyama_2010,ochiai_2011,uppal_2015} all address the
unfolding process. 

Results 
from both unfolding and folding experiments are available, in the case of  the heme-free apomyoglobin 
\cite{griko_1988,jennings_1993,shin_1993,eliezer_1995,eliezer_1996,jamin_1998,jamin_1999,uzawa_2004,nishimura_2006,uzawa_2008,meinhold_2011,nishimura_2011,xu_2012}.  These experiments reveal that unfolding and folding
processes are very similar. 

We have confirmed that in our heating and cooling simulations, 
the unfolding and folding pathways are essentially identical. 

Comparisons of experiments with 
heme containing myoglobin and heme-free apomyoglobin show
that the unfolding proceeds very similarly, in the two cases.  The only real exception is, 
that in the case of apomyoglobin, the F-helix is disordered  at low temperatures \cite{eliezer_1996}.
Accordingly there is no crystallographic data available,  in the case of apomyoglobin,
that we could use to construct a high precision  LGW free energy 
(\ref{E1old}). However, the structural effects of heme during the unfolding process 
are apparently minor. Thus our results are likewise applicable, 
both in the case of heme-free apomyoglobin and heme containing myoglobin.  

%Our experimental reference structure is heme containing myoglobin with 
%PDB code 1ABS \cite{schlichting_1994}, it comes from wild type sperm whale. There are 154 
%amino acids, which are indexed i = 0...153 in the PDB file and in the sequel we follow the PDB indexing.
%The reason why we use 1ABS is, that the thermal  B-factors are very small; the structure has been
%measured at a very low temperature $\sim 20$ K.  This enables us to reach a very high precision, in our
%construction of the LGW free energy.
%

%\vskip cm

\subsection{Multisoliton}

\begin{center}
%\footnotesize
\begin{figure}[h]         
%\begin{center}            
\resizebox{18cm}{!}{\includegraphics[]{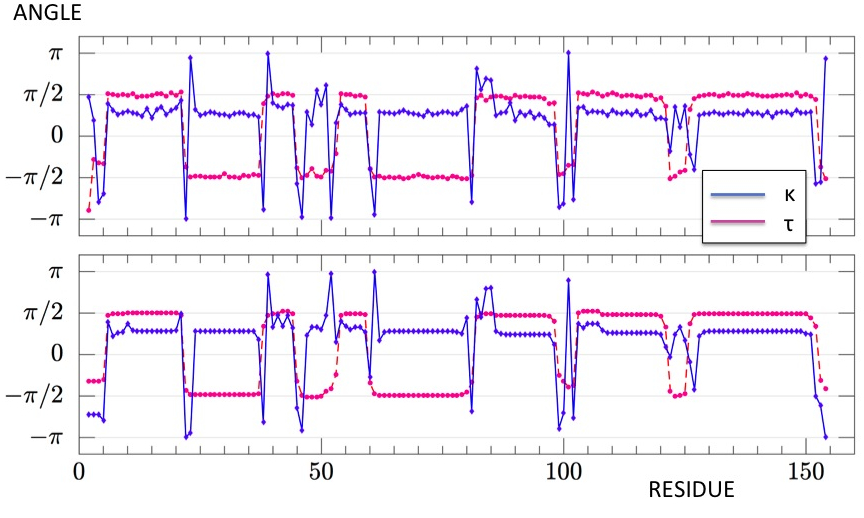}}
\caption{{\it Color online:}  {\it Top:} The bond ($\kappa$) and torsion ($\tau$) angle spectrum of the PDB structure 1ABS. {\it Bottom:} 
The bond ($\kappa$) and torsion ($\tau$) angle spectrum of the 
multisoliton. Note that the angles are defined modulo $2\pi$.}   
\label{fig-5}    
\end{figure}
\end{center}

We start with the construction of the multisoliton solution of (\ref{dnlse}), (\ref{taueq2}),  that 
models the C$\alpha$ backbone of 1ABS \cite{schlichting_1994}.  
We use a combination of the {\tt GaugeIT} and {\tt Propro} packages, described at
\[
{\tt http://folding-protein.org/}
\]
The analysis starts with the inspection of the bond and torsion angle spectrum with the help of
the $\mathbb Z_2$ symmetry (\ref{dsgau}), to identify the individual solitons.
In Figure \ref{fig-5} we show the ($\kappa_i,\tau_i$) spectrum both for 1ABS, and for the multisoliton we have
constructed; the C$\alpha$ RMS distance between the two is around 0.8 \AA.
In Table \ref{param} we show the parameter values that we have found; there are 92 parameters 
that describe the 154 different amino acids. 

\subsection{Stability and reversibility}

We have tested the stability of the multisoliton, by subjecting 
it to repeated heating and cooling simulations using the Glauber algorithm:
We start from the low Monte Carlo temperature factor value $T = 10^{-17}$ where we observe no thermal 
fluctuations. We increase the temperature factor linearly on a logarithmic scale, so that we reach the 
value $T=10^{-4}$ after 5 million MC steps.  We then fully thermalise the configuration at 
$T = 10^{-4}$, during another
5 million MC steps.  Finally, we cool it down, back to  the original low temperature value, during 5 million
steps.

%%%%%%%%%%%%%%%%%%%%%%%%%%%%%%%%%%%%%%%%%%%%
%
%
%
%
%
%                           FIGURE 5
%
%
%
%
%
% 
%%%%%%%%%%%%%
%%%%%%%%%%%%%
%%%%%%%%%%%%%%
\onecolumngrid

%\onecolumngrid
\begin{center}
\begin{table}[h!]
 \begin{tabular} {|l | l | l | l | l | l | l | l | l | }
\hline
Index & c1 & c2 & m1 & m2 & b & d & e & q\\ \hline
1 & 12.0778 & 3.91722 & 1.01149 & 1.54173 & 6.2835e-08 & 3.1072e-08 & 4.1929e-08 & -2.17006e-06\\ \hline
2 & 3.43595 & 2.029 & 1.58004 & 1.51381 & 1.00719e-08 & 7.23577e-08 & 1.2505e-08 & 1.08099e-06\\ \hline
3 & 7.31917 & 0.814575 & 1.50642 & 1.54302 & 1.84985e-09 & 1.01644e-07 & 2.70434e-10 & -4.82807e-08\\ \hline
4 & 2.13766 & 0.656997 & 1.65579 & 1.60224 & 2.87694e-09 & 9.05135e-08 & 2.55083e-11 & -1.20232e-06\\ \hline
5 & 0.88539 & 5.97185 & 1.36452 & 1.53686 & 3.83823e-09 & 2.34144e-07 & 1.18097e-08 & 3.30105e-07\\ \hline
6 & 8.71177 & 0.83374 & 1.55042 & 1.53703 & 2.43781e-09 & 9.64278e-08 & 5.11205e-11 & -4.778e-07\\ \hline
7 & 0.97324 & 2.14009 & 1.46169 & 1.54621 & 9.51103e-15 & 7.4009e-09 & 3.47317e-10 & -3.83551e-09\\ \hline
8 & 1.32577 & 2.91054 & 1.47714 & 1.01994 & 2.72523e-14 & 1.37454e-13 & 1.74597e-14 & -5.60294e-13\\ \hline
9 & 10.4862 & 4.24384 & 1.22245 & 1.65318 & 6.12822e-09 & 1.21357e-07 & 4.95752e-11 & -1.37175e-06\\ \hline
10 & 0.800415 & 1.28973 & 1.5154 & 1.60278 & 3.91353e-08 & 2.03487e-07 & 7.30035e-12 & -1.13574e-06\\ \hline
11 & 3.15255 & 0.914751 & 1.55827 & 1.55151 & 3.86819e-09 & 1.07811e-07 & 3.74786e-11 & -1.02768e-06\\ \hline
12 & 1.0122 & 1.06369 & 1.40009 & 1.32823 & 5.68916e-09 & 1.11761e-07 & 2.19282e-10 & -8.62094e-07\\ \hline
\end{tabular}
\caption{The parameters in the energy function for 1ABS. The ensuing profile is shown in Figure \ref{fig-5} bottom. 
The first column is the index of the individual solitons; there are a total of ten individual
soliton profiles along the entire myoglobin backbone. In each soliton, we divide 
the parameters $c$ and $m$ into $c_1, c_2$ and $m_1, m_2$ to reflect the asymmetry of the soliton around its center.}
\label{param}
\end{table}
\end{center}
%\end{landscape}
%\twocolumngrid
%
%
%
%
%
\twocolumngrid
%%%%%%%%%%%%%
%%%%%%%%%%%%%
%%%%%%%%%%%%%%
%%%%%%%%%%%%%
%%%%%%%%%%%%%
%%%%%%%%%%%%%%
%
%
%
%\begin{landscape}
%%%%%%%%%%%%%
%%%%%%%%%%%%%
%%%%%%%%%%%%%%

For production, we have performed 100 full heating-cooling cycles.  
The Figure \ref{fig-6} shows the evolution of RMS distance to the low temperature multisoliton, during the
heating and cooling cycle; we find that the heating and cooling proceed very symmetrically, and the configuration
returns to the original  low temperature structure at the end of the cycle.
%
%
%
%%%%%%%%%%%%%%%%%%%%%%%%%%%%%%%%%%%%%%%%%%%%
%
%
%
%
%
%                           FIGURE 6
%
%
%
%
%
% 
%%%%%%%%%%%%%
%%%%%%%%%%%%%
%%%%%%%%%%%%%%
{
\footnotesize
\begin{figure}[h]         
\centering           
  \resizebox{8 cm}{!}{\includegraphics[]{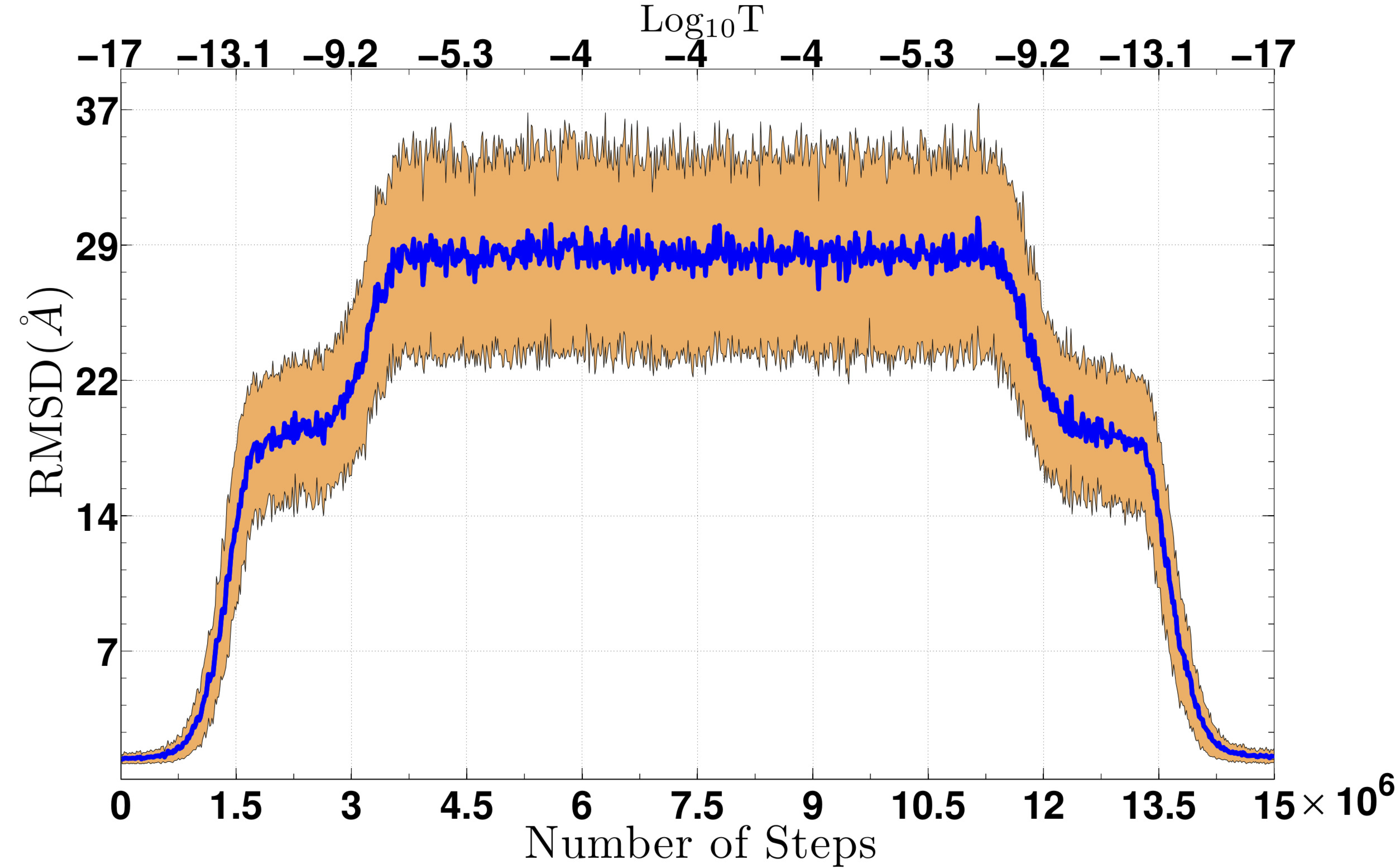}}
\caption {{\it Color online:} Evolution of RMS distance between the 1ABS and the multisoliton, during the heating-cooling cycle.
No further increase in the RMS distance is observed, if the Monte Carlo temperature factor is further increased.  The blue line is 
the average and the orange band displays the one standard deviation from the average.}   
\label{fig-6}    
\end{figure}
}
%%%%%%%%%%%%%
%%%%%%%%%%%%%
%%%%%%%%%%%%%%
%%%%%%%%%%%%%
%%%%%%%%%%%%%
%%%%%%%%%%%%%%
%
%
%

\subsection{Heating myoglobin}

In our production runs for heating simulations, described in the sequel, 
we have increased the temperature factor from $T=10^{-17}$ to $T=10^{-4}$ during 6 million Monte Carlo steps; 
the results do not depend on the number of steps, as long as this number is not very small.  We have 
performed 100 independent full length heating simulations. 
We have in particular confirmed that the 
heating process is fully reversible: Upon cooling the system from the high temperature value back to the original low 
temperature value the structure folds back to the native conformation. The unfolding/folding pathways are essentially
identical.

Our  simulations are extremely time efficient. For example,  with a MacPro workstation 
a complete  heating and cooling cycle takes 
around ten seconds of {\it in silico} time, in a single processor. By comparison, the experimentally 
observed folding time of apomyoglobin is around 2.5 seconds \cite{jennings_1993}.

\subsection{$\alpha$-helical content}

The 
monitoring of $\alpha$-helical content using {\it e.g.} circular dichroism (CD) spectroscopy  gives 
an indication how the unfolding proceeds.  Accordingly we have estimated the $\alpha$-helical content
during our heating simulations.

We define a  C$\alpha$ atom which is centered at $\mathbf r_{i}$, 
to be in an $\alpha$-helical position when  
$| \mathbf r_{i+4} - \mathbf r_{i} |\approx  6.2 \ \pm 0.5$ \AA ~ and $|\tau_i - \tau_{0}| < 0.6 $ (rad) 
where $\tau_{0}$ is the experimental average value of the $\alpha$-helical torsion angle; 
we have deduced these values from a statistical analysis of PDB structures.

We deduce from Figure \ref{fig-5} that folded
myoglobin  has a substantial $\alpha$-helical content: In 1ABS, around $\sim 72 \%$ of
the  C$\alpha$ atoms are in $\alpha$-helical position, and there is also a 
small fraction in the closely related $3/10$ position.

We have investigated how the $\alpha$-helical content depends on the Glauber temperature factor $T$,
during the unfolding process. The results, shown in Figure \ref{fig-7},  are qualitatively very 
similar to the  experimentally observed  circular dichroism data  shown in  Figure 2 of reference 
\cite{moriyama_2010}. We note that according to \cite{moriyama_2010} the heme becomes 
irreversibly damaged  at around  $\sim 75-80^{\rm \ o}$C (the red dashed line in Figure \ref{fig-7}).  
However, we point out that the structural stability of myoglobin varies between species;
the myoglobin in \cite{moriyama_2010} 
is from horse heart.  

%
%
%
%%%%%%%%%%%%%%%%%%%%%%%%%%%%%%%%%%%%%%%%%%%%
%
%
%
%
%
%                           FIGURE 7
%
%
%
%
%
% 
%%%%%%%%%%%%%
%%%%%%%%%%%%%
%%%%%%%%%%%%%%
{
\footnotesize
\begin{figure}[h]         
\centering       
  \resizebox{8 cm}{!}{\includegraphics[]{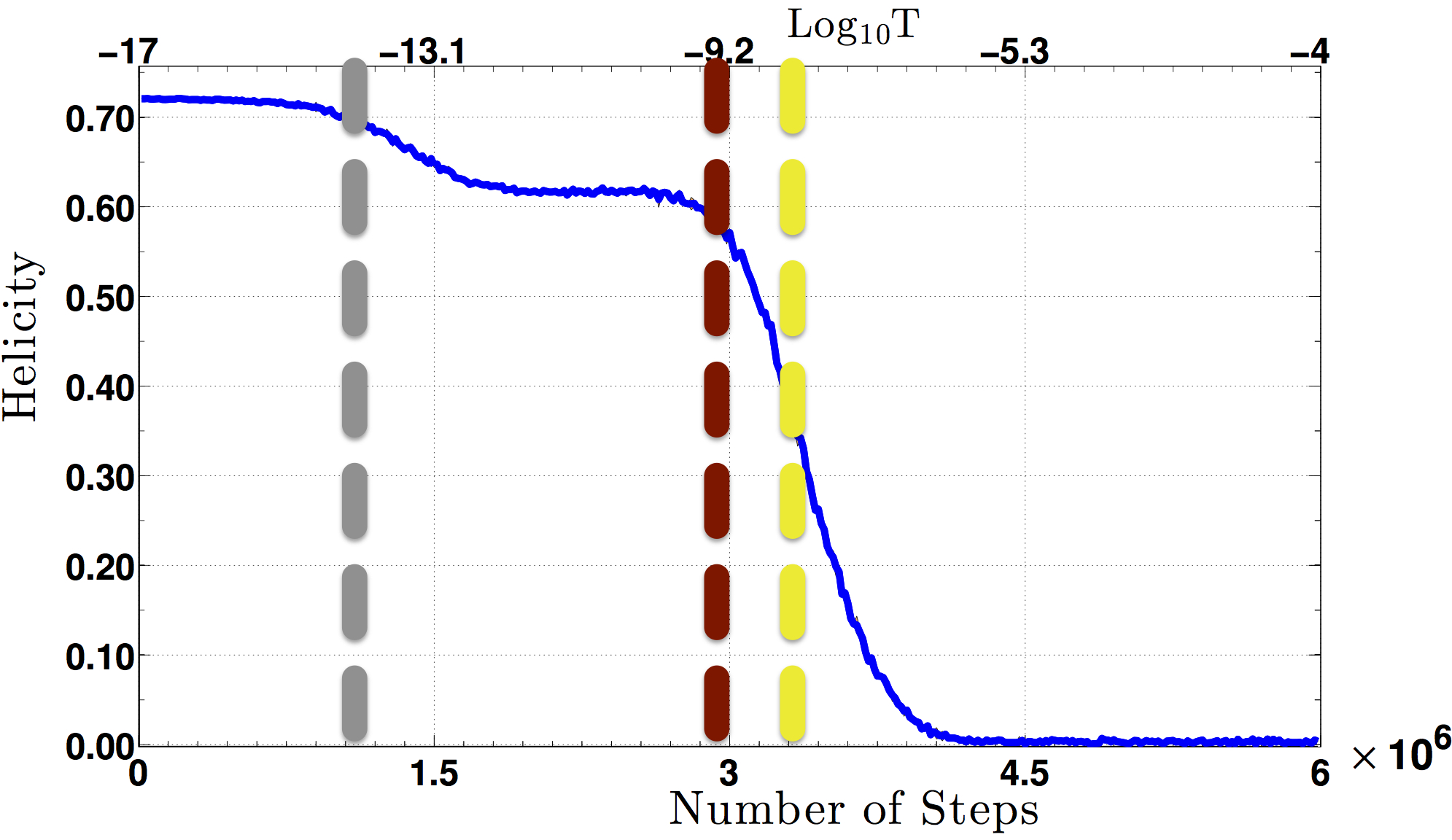}}
\caption {{\it Color online:} Simulated $\alpha$-helical content (in $\%$) as a function of Glauber  temperature $T$: The dashed grey line estimates $\sim 25^{\rm o}C$, the dashed red line estimates $
        \sim 75^{\rm o}C$
        and the dashed yellow line estimates $\sim90^{\rm o}C$ in Figure 2 of  \cite{moriyama_2010}
        }   
\label{fig-7}    
\end{figure}
}
%%%%%%%%%%%%%
%%%%%%%%%%%%%
%%%%%%%%%%%%%%
%%%%%%%%%%%%%
%%%%%%%%%%%%%
%%%%%%%%%%%%%%
%
%
%

\subsection{Radius of gyration and its susceptibility}

In protein unfolding experiments, the radius of gyration $R_g$ and its evolution is used widely to monitor  
the progress. In Figure \ref{fig-8} 
we show how the radius of gyration evolves during our heating (unfolding) simulations. 
%
%
%
%%%%%%%%%%%%%%%%%%%%%%%%%%%%%%%%%%%%%%%%%%%%
%
%
%
%
%
%                           FIGURE 8
%
%
%
%
%
% 
%%%%%%%%%%%%%
%%%%%%%%%%%%%
%%%%%%%%%%%%%%
{
\footnotesize
\begin{figure}[h]         
\centering           
  \resizebox{8.5 cm}{!}{\includegraphics[]{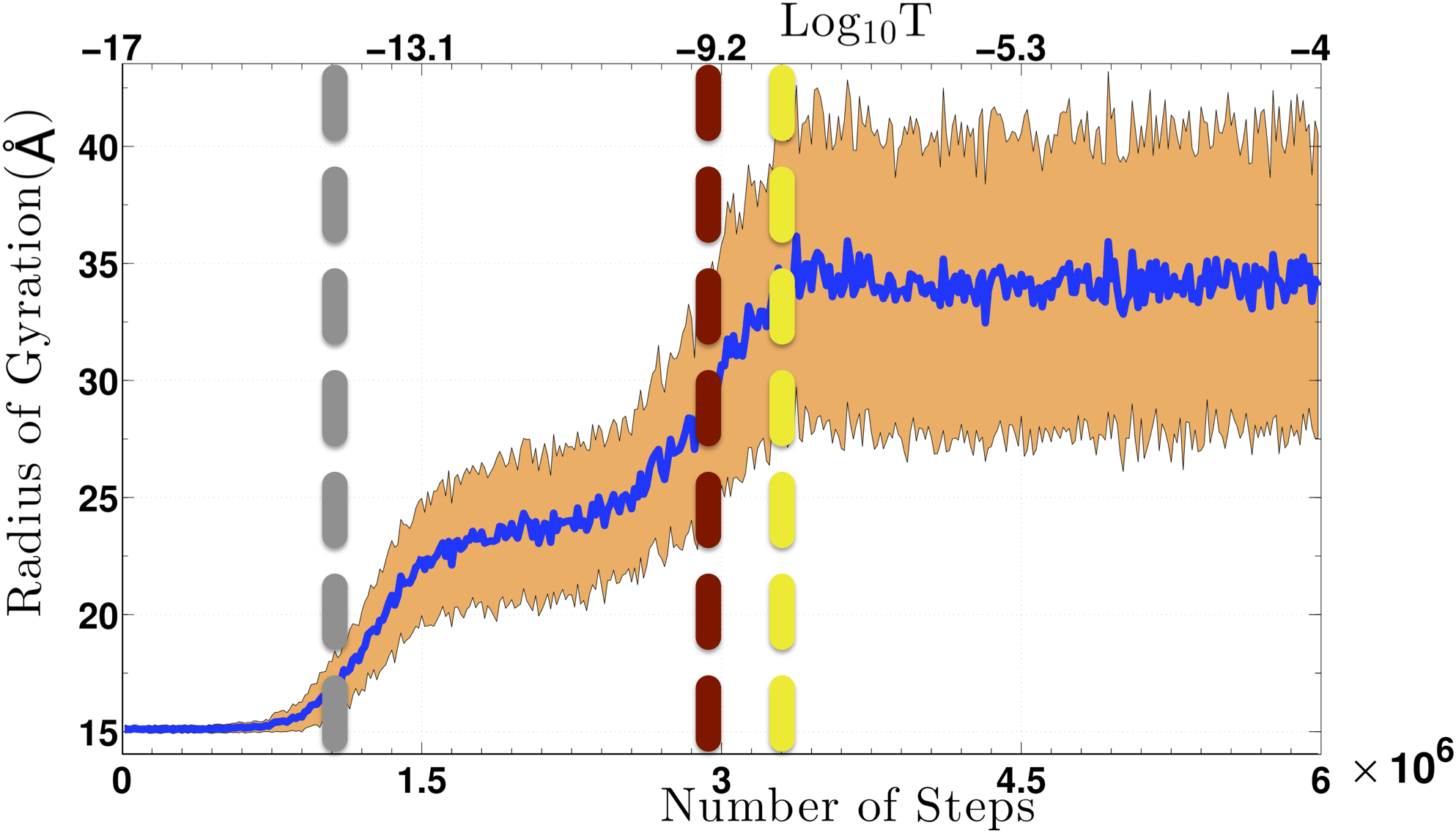}}
\caption {{\it Color online:} The dependence of radius of gyration as a function of Glauber temperature factor. As in Figure  \ref{fig-6},
the blue line is average value 
and the orange band denotes the one standard deviations fluctuation distance.
A comparison shows 
that the transition temperatures of $R_g$ are slightly lower  than in the case of RMSD, in Figure \ref{fig-6}. 
The dashed grey, red and yellow lines are as in Figure \ref{fig-7}.}   
\label{fig-8}    
\end{figure}
}
%%%%%%%%%%%%%
%%%%%%%%%%%%%
%%%%%%%%%%%%%%
%%%%%%%%%%%%%
%%%%%%%%%%%%%
%%%%%%%%%%%%%%
%
%
%
We observe the presence of a folding intermediate between $\log_{10} T_L \approx  -12.8$
and $\log_{10} T_H \approx -9.5$. 
The ensemble average value $R_g \approx \ 24$ \AA ~ of the folding
intermediate is very close to the experimentally observed value $R_{exp} \sim \ 23.6$ \AA~ 
of the molten globule, measured in the case of the apomyoglobin  \cite{eliezer_1995,nishimura_2006}. 
The increase in $R_g$ during the first transition in Figure \ref{fig-8}, 
from native state to molten globule, is around 9 \AA. 
This is larger than the experimentally observed 1-7 \AA~ low pH values in the 
apomyoglobin  \cite{eliezer_1995,nishimura_2006}, but the difference is in line with the 
observation that at low temperatures the radius of gyration of apomyoglobin is larger 
than that of heme containing myoglobin \cite{eliezer_1995,nishimura_2006}. 
 Between the molten globule and the fully unfolded state the ensemble average difference
$ R_g \sim 10$ \AA ~ that we find, is very close to the experimentally measured $11 \pm 2$ \AA~  
low pH value, in the case of apomyoglobin \cite{eliezer_1995}. 

The transition temperatures during the unfolding process can be estimated
by evaluating the following {\it radius of gyration susceptibility}
\begin{equation}
\mathcal X_{g}(T) \ = \   \frac{d R_{g} (T) }
{d \log_{10} T}
\la{logRg}
\end{equation}
To evaluate this quantity,  we introduce a  fitting  procedure 
where we first approximate $R_{g}(T)$  by a  function of the form
\begin{equation*}
\log_{10} (R_{g}(T) ) \ \approx \  R_{g}^{\mathrm{fit}}(\log_{10} T)
\end{equation*}
where we choose
\[
R_{g}^{\mathrm{fit}}(x)  \  = \  h_1 + h_2 \arctan[h_3 (x - x_1)]
\]
\begin{equation}
+ h_4  \,x \arctan[h_5 (x - x_2)] - h_6 x\,.
\la{eq:E:fit2}
\end{equation}
This function form has been introduced and utilised in \cite{chernodub_2011}, in a related context.
The numerical values of the parameters  $h_1 \dots h_6$ and $x_{1,2}$ are determined by a fit to the  numerical
values of $R_{g}(T)$ (not shown here).  We then use  the peaks in (\ref{logRg}) to 
determine the transition values of $T$.  There are two peaks during the heating process, at MC temperature factor
values 
\begin{equation}
\begin{matrix}
T_1 \approx 10^{-8.6} \\
T_2 \approx 10^{-13.6}
\end{matrix}
\label{T12}
\end{equation}

\vskip 0.3cm

We conclude that both in terms of $\alpha$-helical content and radius of gyration, our model appears to
correctly describe  the observed 
reversible myoglobin unfolding dynamics  below $\log_{10}  T_c \approx -8.0$.  
In particular, in Figure \ref{fig-8} we observe a folding intermediate between  
$\log_{10}  T_L \approx -12.8$ and $ \log_{10}  T_H \approx -9.5$. These three temperature factor values
correspond to the  three dashed lines (grey, red, yellow) 
which we have also  identified in  Figure \ref{fig-7}.

\subsection{Energy susceptibility}

The various transitions can also be monitored by changes in energy, 
in terms of {\it energy susceptibility}. For this we first evaluate  
the average internal energy $<\!E\!>$ as a function of the temperature factor $T$, in thermal equilibrium,
using the LGW Hamiltonian (\ref{E1old})
\[
<\!E\!> \ = \ -  \frac{\partial}{\partial \beta} Tr \exp \{ -\beta H \}  \ \ \ \ \ \  ( \beta = \frac{1}{T} )
\] 
The energy susceptibility is akin  to the heat 
capacity,
\begin{equation}
\chi_E = \frac{ d <\! E \!> }{ d\log_{10}(T) }
\la{heatc}
\end{equation}

We use a fitting function such as  (\ref{eq:E:fit2}), to numerically estimate (\ref{heatc}).  
In Figure \ref{fig-9} we display the thermal equilibrium state energy susceptibility, that we have computed.
We observe in Figure \ref{fig-9} (A) a clear peak,  in the high temperature regime, at
\[
\log_{10} T_3\approx -0.85
\]
It appears that, thus far, this peak has not been observed experimentally; the corresponding physical 
temperature value appears to be quite high. 
%
%
%
%
%
%%%%%%%%%%%%%%%%%%%%%%%%%%%%%%%%%%%%%%%%%%%%
%
%
%
%
%
%                           FIGURE 9
%
%
%
%
%
% 
%%%%%%%%%%%%%
%%%%%%%%%%%%%
%%%%%%%%%%%%%%
{
\footnotesize
\begin{figure}[h]         
\centering            
  \resizebox{9 cm}{!}{\includegraphics[]{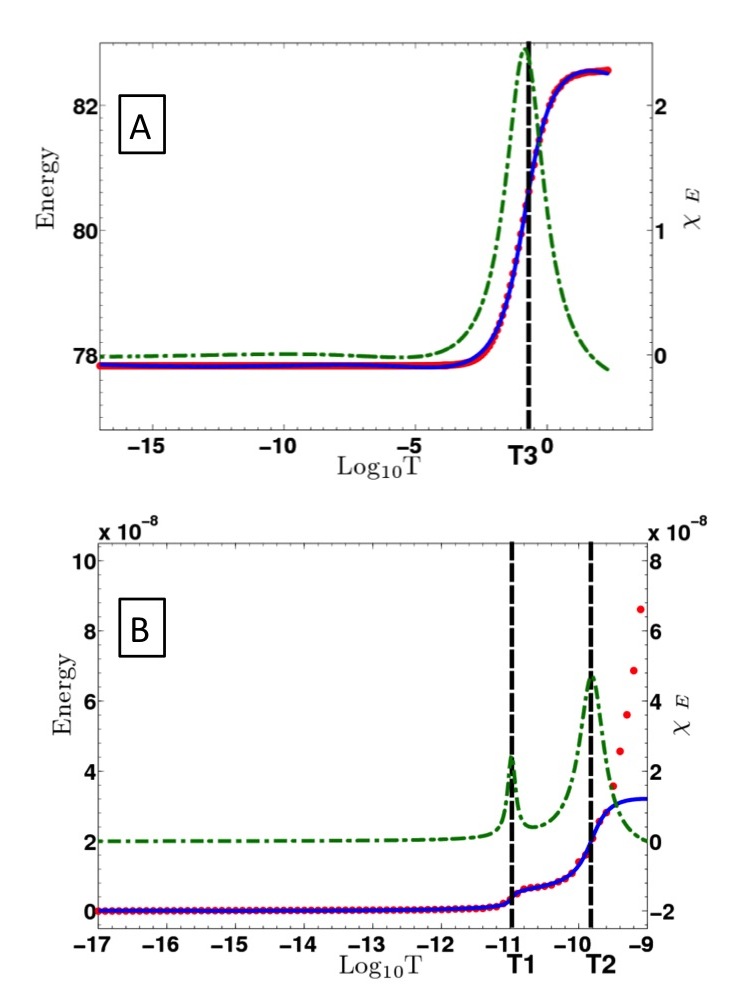}}
\caption {{\it Color online:} The average energy and the energy susceptibility plots for different temperature ranges. 
(A) $T \in [10^{-17},10^{4}]$. (B) energy and heat capacity zoomed in $T \in [10^{-17},10^{-9}]$, where 
the scale on the energy-axis has been subtracted by the energy at native state (equaling to $\sim$77.83 in units
of \ref{E1old})).
The red solid dots are the average energy 
values at corresponding temperature, the blue line is the energy curve fittings based on  
(\ref{eq:E:fit2}). The green dashed line is the energy susceptibility calculated from (\ref{heatc}). 
The temperatures of the energy susceptibility peaks are denoted as T1,T2,T3. 
}   
\label{fig-9}    
\end{figure}
}
%%%%%%%%%%%%%
%%%%%%%%%%%%%
%%%%%%%%%%%%%%
%%%%%%%%%%%%%
%%%%%%%%%%%%%
%%%%%%%%%%%%%%
%
%
%
We also observe two peaks at  lower temperature values
\[
\begin{matrix} 
\log_{10}T_1\approx -11.0 \\
\log_{10}T_2 \approx -9.8
\end{matrix}
\]
See Figure \ref{fig-9} (B). Note that in Figure \ref{fig-9} (A) these two peaks are not very visible, 
as their heights are much lower than the height of the high temperature peak. The normalisation of the 
peak height reflects the relation (\ref{Ttheta}).

Our observation of two lower temperature peaks, very close to each other, is consistent 
with the presence of an experimentally measured single wide peak \cite{griko_1988} where the 
experimentally heat capacity peak at pH 5.0 is broad; see Figure 5 in  \cite{griko_1988}.

\subsection{$\alpha$-helix de-nucleation}

We monitor details of the unfolding process, by evaluating  the temperature dependence in the 
fluctuations $\Delta \tau_i$ of the individual backbone  torsion angles, defined as follows
\begin{equation}
\Delta\tau_i \ = \ \sqrt{ \frac{1}{N}  \sum_{k=1}^{N} (\tau_{i,k} - \bar\tau_i)^2 }
\label{taui}
\end{equation}
The index $k$ counts the conformation in a given heating simulation, and the average is
over the entire ensemble of $N=100$ heating simulations. 
In Figures \ref{fig-10a} and \ref{fig-10b} we display the evolution of (\ref{taui}), at six dffferent temperatures during 
the unfolding process.
%
%
%
%
%
%%%%%%%%%%%%%%%%%%%%%%%%%%%%%%%%%%%%%%%%%%%%
%
%
%
%
%
%                           FIGURE 10
%
%
%
%
%
% 
%%%%%%%%%%%%%
%%%%%%%%%%%%%
%%%%%%%%%%%%%%
{
\footnotesize
\begin{figure}[h]         
\centering         
  \resizebox{9.3 cm}{!}{\includegraphics[]{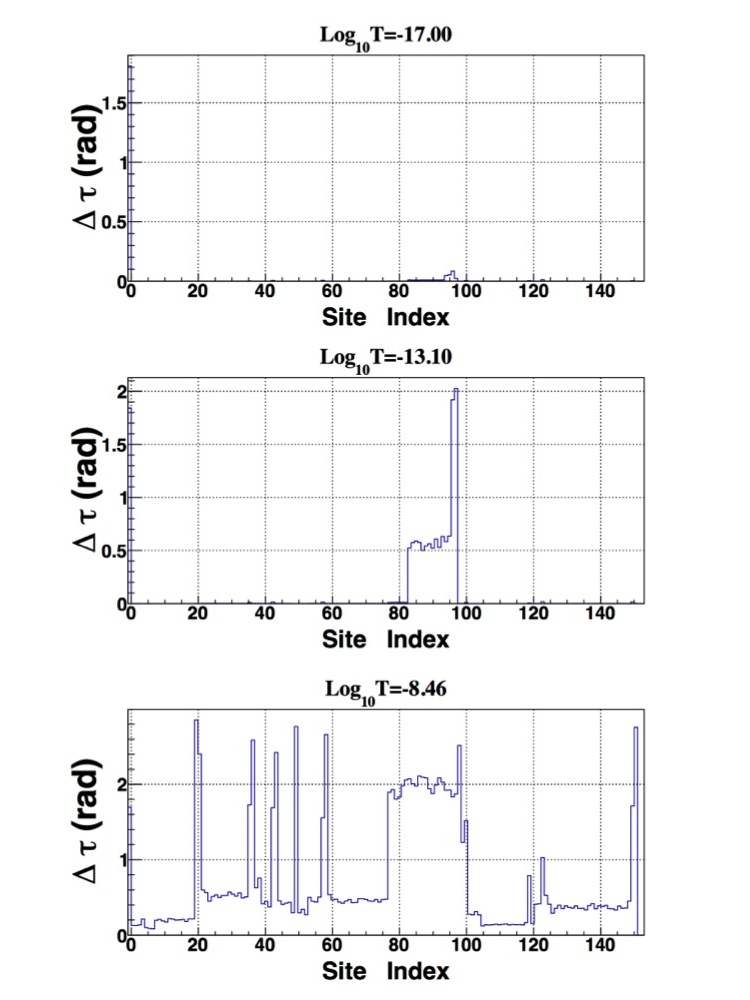}}
\caption { {\it Color online:} 
The average values (\ref{taui}) at three different temperature factor values $\log_{10}(T) = -17, \  
-13.1, \  -8.46$.
}   
\label{fig-10a}    
\end{figure}
}
%
%
%
%
%
%%%%%%%%%%%%%%%%%%%%%%%%%%%%%%%%%%%%%%%%%%%%
%
%
%
%
%
%                           FIGURE 11
%
%
%
%
%
% 
%%%%%%%%%%%%%
%%%%%%%%%%%%%
%%%%%%%%%%%%%%
{
\footnotesize
\begin{figure}[h]         
\centering            
  \resizebox{9.3cm}{!}{\includegraphics[]{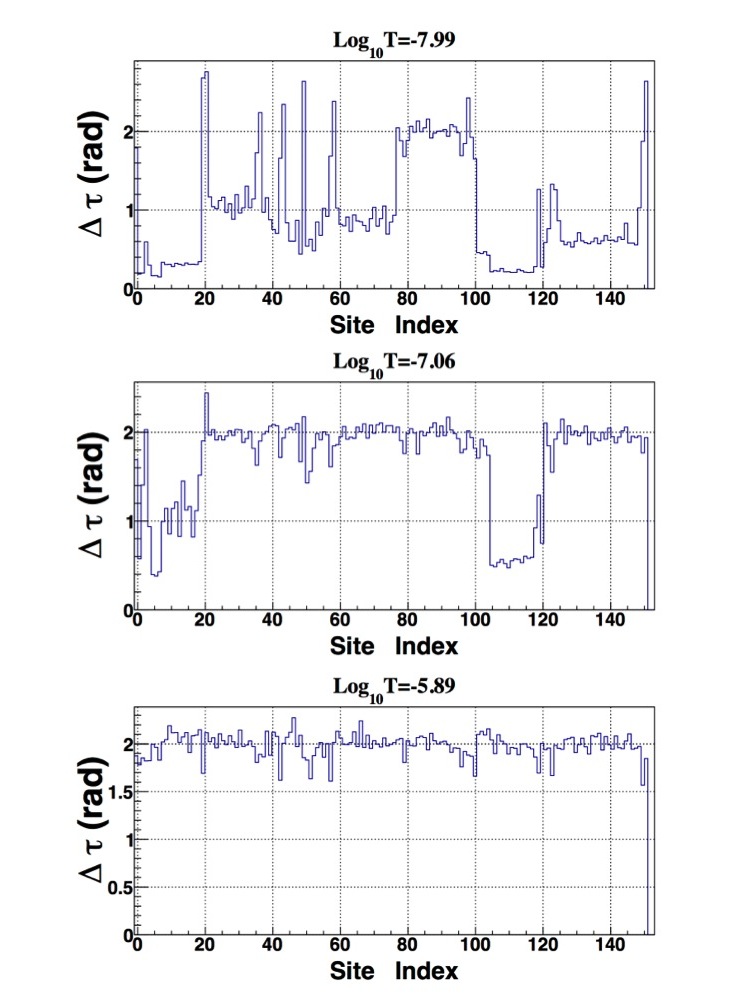}}
\caption { {\it Color online:} 
The average values (\ref{taui}) at three different temperature factor values $\log_{10}(T) = -7.99, \ -7.06, \ -5.89$.
}   
\label{fig-10b}    
\end{figure}
}
We observe in particular, how different helices become de-nucleated at different temperatures, during the unfolding process. 

To determine the de-nucleation temperatures of the eight individual  helical segments 
$X = (A,B, \dots ,H)$ in the natively folded myoglobin, we evaluate the following average values of 
the $\tau$-fluctuations,
\begin{equation}
\Delta \tau_X \ = \sqrt{ \frac{1}{|X|}  \sum_{X} (\Delta \tau_i)^2 }
\label{taux}
\end{equation}
Here $|X|$  is the number of residues in the native helical segment $X$, and
both $\Delta \tau_i$ and $\Delta \tau_X$ 
are evaluated  at 1000 different sampling temperatures, during the heating. 

We introduce a susceptibility akin (\ref{logRg}), (\ref{heatc})  to monitor the individual  
$\alpha$-helix unfolding
\begin{equation}
\mathcal X_\tau \ = \ \frac{ d \Delta \tau_X}{d \log_{10} T}
\label{Xtau}
\end{equation}
Figure \ref{fig-12} summarises the results obtained from equations (\ref{logRg}), (\ref{heatc}) and (\ref{Xtau}).
%
%
%
%
%
%%%%%%%%%%%%%%%%%%%%%%%%%%%%%%%%%%%%%%%%%%%%
%
%
%
%
%
%                           FIGURE 12
%
%
%
%
%
% 
%%%%%%%%%%%%%
%%%%%%%%%%%%%
%%%%%%%%%%%%%%
{
\footnotesize
\begin{figure}[h]         
\centering            
  \resizebox{8cm}{!}{\includegraphics[]{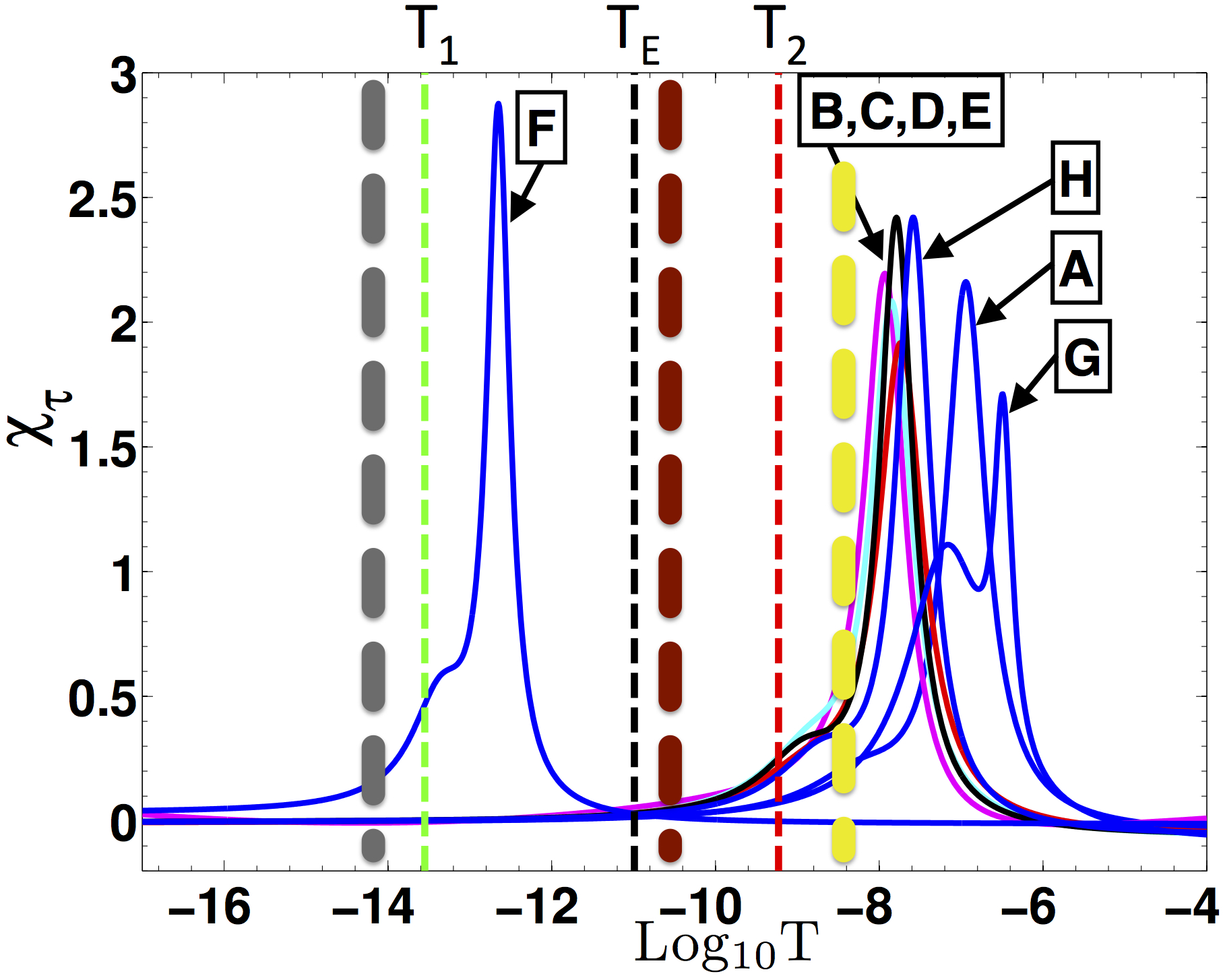}}
\caption { {\it Color online:} 
The susceptibility (\ref{Xtau})  in myoglobin. Helices B,C,D,E are identified by the colors magenta, cyan, red, 
black, respectively. The vertical  black dotted line denotes the maximum of (\ref{heatc}) at  $\log_{10}T_E \approx - 10.2$. 
The vertical green and red dotted lines denote two $R_g$ susceptibility peaks at $\log_{10}T_1 \approx - 13.8$ and 
at $\log_{10}T_2 \approx - 9.3$.   The dashed grey, red and yellow lines are as in Figures \ref{fig-7} and \ref{fig-8}.  Units 
along ordinate  
 derive  from (\ref{Xtau}).
}   
\label{fig-12}    
\end{figure}
}

\subsection{Contact maps}

We monitor long range interactions  between any pair of different backbone segments, in terms of a  contact map. 
For this we denote by
 \[
 d_{ij} = |\mathbf r_i - \mathbf r_j | 
 \]
the distance between 
{\it any} two C$\alpha$ atoms.  We define a scoring function for each pair
of  helical segments $X,Y = (A,B, \dots ,H) $ in the natively folded myoglobin as follows,
\begin{equation}
S_{ij} \ = \ \left\{ \begin{matrix} 0 & ~ {\rm for } ~ & d_{ij} > 12 \\ 
\frac{ 12 - d_{ij} }{4} & ~ {\rm for } ~ & 8 \leq d_{ij} \leq  12 \\
1 & ~ {\rm for } ~ & d_{ij} < 8
\end{matrix} \right.
\la{S}
\end{equation} 
We define the average contact by
\begin{equation}
\mathcal S_{X,Y} = \frac{ \sum_{i\in X} \sum_{j\in Y} \overline{ S_{ij} - S_{ij0} }}{\rm{min}(|X|,|Y|)}
\la{Save}
\end{equation}
Here $|X|, \ |Y|$ are the lengths of the helical segments $X,\ Y$, and $S_{ij0}$ is the value of 
(\ref{S}) in the native state, and the average is taken over the entire ensemble.
A detailed analysis of (\ref{taui})-(\ref{Save}) confirms that in our simulations  the
thermal unfolding does indeed proceed sequentially, through helical intermediates, in a manner 
which is fully in line with the experimental observations: The individual contact maps for each of the eight helices
are presented in Figures \ref{fig-13}-\ref{fig-16}.  
%
%
%
%
%
%%%%%%%%%%%%%%%%%%%%%%%%%%%%%%%%%%%%%%%%%%%%
%
%
%
%
%
%                           FIGURE 13
%
%
%
%
%
% 
%%%%%%%%%%%%%
%%%%%%%%%%%%%
%%%%%%%%%%%%%%
{
\footnotesize
\begin{figure}[t]         
\centering            
  \resizebox{7.5cm}{!}{\includegraphics[]{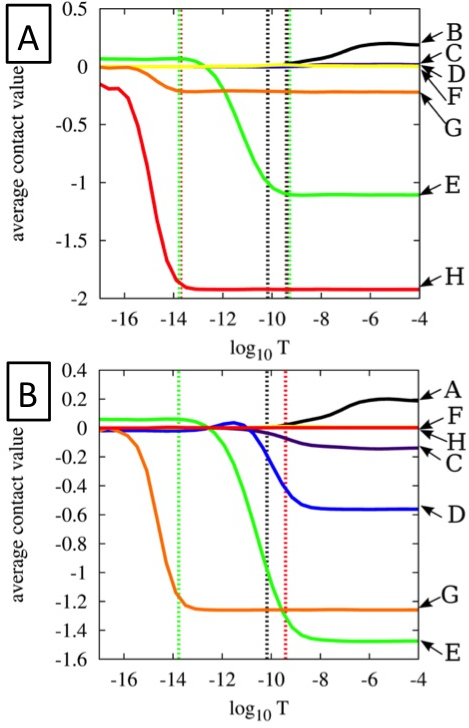}}
\caption { {\it Color online:} {\tt Top:} Contact maps between helix A  with all other helices. {\tt Bottom:} 
Contact maps between helix B with all other helices. 
The value of average contacts is defined by (\ref{Save}). The vertical lines 
indicate major  changes in $R_g$ and energy:   
The vertical red 
dashed line denotes the high temperature peak of the susceptibility (\ref{logRg}) and the vertical green dashed 
line denotes its  low temperature peak.  The vertical black dashed line denotes low 
temperature energy susceptibility peak. 
}   
\label{fig-13}    
\end{figure}
}
%
%
%
%
%
%
%
%

%
%
%%%%%%%%%%%%%%%%%%%%%%%%%%%%%%%%%%%%%%%%%%%%
%
%
%
%
%
%                           FIGURE 14
%
%
%
%
%
% 
%%%%%%%%%%%%%
%%%%%%%%%%%%%
%%%%%%%%%%%%%%
{
\footnotesize
\begin{figure}[b]         
\centering          
  \resizebox{7.5cm}{!}{\includegraphics[]{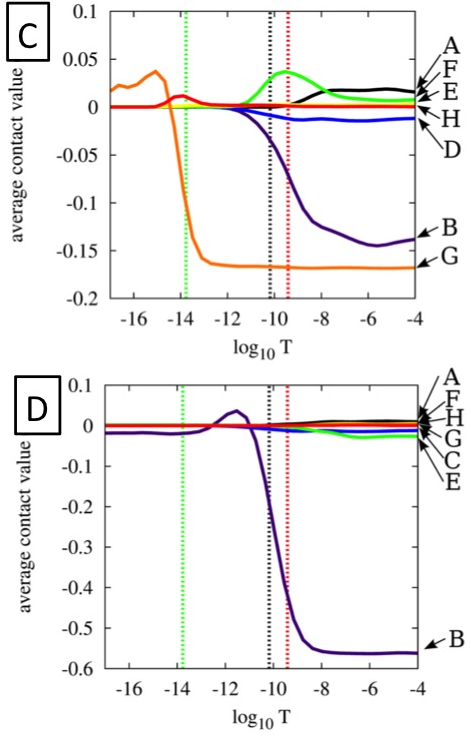}}
\caption { {\it Color online:} Same as in Figure \ref{fig-13},  for helices C and D.
}   
\label{fig-14}    
\end{figure}
}
%
%
%
%
%
%

%
%
%
%
%
%
%
%
%
%%%%%%%%%%%%%%%%%%%%%%%%%%%%%%%%%%%%%%%%%%%%
%
%
%
%
%
%                           FIGURE 15
%
%
%
%
%
% 
%%%%%%%%%%%%%
%%%%%%%%%%%%%
%%%%%%%%%%%%%%
{
\footnotesize
\begin{figure}[t]         
\centering           
  \resizebox{7.5cm}{!}{\includegraphics[]{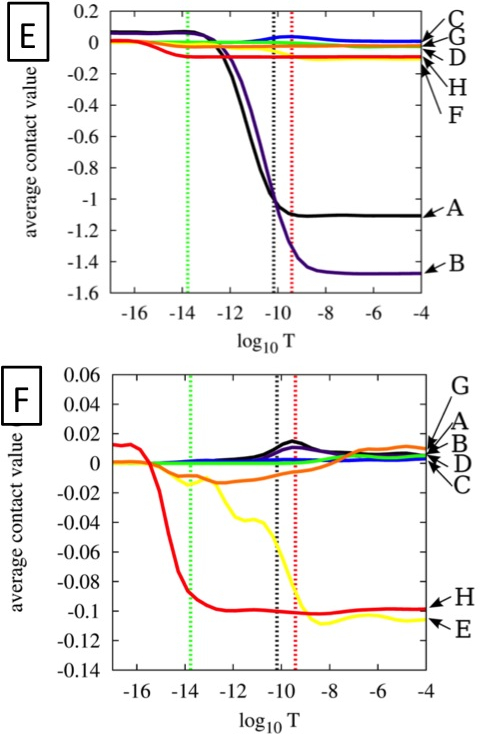}}
\caption { {\it Color online:} Same as in Figure \ref{fig-13} for helices E and F.
}   
\label{fig-15}    
\end{figure}
}
%
%
%
%
%
%
%

%
%
%
%
%
%
%
%
%%%%%%%%%%%%%%%%%%%%%%%%%%%%%%%%%%%%%%%%%%%%
%
%
%
%
%
%                           FIGURE 16
%
%
%
%
%
% 
%%%%%%%%%%%%%
%%%%%%%%%%%%%
%%%%%%%%%%%%%%
{
\footnotesize
\begin{figure}[h]         
\centering           
  \resizebox{7.5cm}{!}{\includegraphics[]{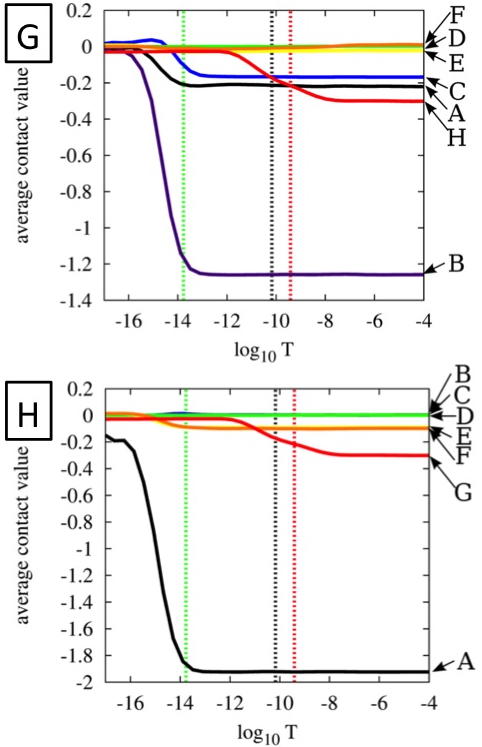}}
\caption { {\it Color online:} Same as in Figure \ref{fig-13} for helices G and H.
}   
\label{fig-16}    
\end{figure}
}

From the contact maps shown in Figures \ref{fig-13}-\ref{fig-16}  
we estimate the critical temperature values, where 
a long range interaction between each pair of helices disappears. The critical values of 
temperature factors $T$ are shown in Table \ref{table_2}, for each pair of helices.

\subsection{A summary of observations}

According to our simulations,  the
$\alpha$-helix unfolding in myoglobin 
takes place {\it only} in conjunction with the disappearance of 
long range interactions between the ensuing helix and another helical segment of the backbone. In particular,
we observe that 
the F helix becomes unstable already at relatively low temperatures, before the molten globule forms.
This is consistent with experiments that are made with heme containing myoglobin, and  in line with the 
observed disordered character of the F 
helix, in the case of apomyoglobin.  We find that
the helices B, C, D and E remain largely stable during the molten globule phase.
Their unfolding coincides with the melting of the molten globule, apparently
in conjunction with the irreversible destruction of the heme pocket \cite{moriyama_2010}.
Subsequently we observe the unfolding of A and H helices, in line  
with the original folding prediction in \cite{matheson_1978,gerritsen_1985}. In our simulations,
the G helix is the last to unfold.

The following diagram summaries our observed thermal unfolding of the myoglobin,
with increasing temperature: 

\begin{equation}
F  \longrightarrow {\rm molten \ globule} \longrightarrow  B,C,D,E \longrightarrow  A,H \longrightarrow G
\label{order}
\end{equation} 

When all helices have become unfolded, which occurs at around $\log_{10}T \approx -6$ (yellow line in Figures \ref{fig-7},
\ref{fig-8}), 
the backbone appears to be in the universality class of self-avoiding random  walk. 
In particular, we observe no further increase in 
$R_g$ even when the ambient temperature becomes substantially increased .

When we adiabatically cool the system down to the original temperature,  we observe that the helices 
form in an order which is opposite  to that during the heating process.

\section{Conclusions}

In conclusion, we have combined the Landau-Ginzburg-Wilson approach
with non-equilibrium Glauber dynamics, to model the way how 
myoglobin unfolds when the ambient temperature increases. 
All our simulation results appear to be in excellent 
agreement with available experimental results. In addition, we have 
proposed new observables including a detailed contact 
map between different helical segments. This  could be tested in future experiments,
to estimate the range of validity of the LGW approach in the case of proteins.

The approach that we have developed models both the natively folded low temperature 
structure and its thermally driven unfolding process,  in terms of a multi-soliton solution
of the pertinent  Landau-Ginzburg-Wilson energy function and its Glauber  dynamics.  In particular, 
the multi-soliton is a solution of a universal discrete nonlinear Schr\"odinger equation. It  
approximates the C$\alpha$ backbone profile in the limit where
the spatial variations along the backbone have a long wavelength.  This is the limit, where one generally expects
the Landau-Ginzburg-Wilson approach to become valid.  Moreover, the presence of
solitons furnishes the energy function with a substantial predictive 
power: The number of free parameters is even much less than the number of amino acids in myoglobin.
 
Our results  propose  that the unfolding process of a myoglobin is primarily driven by 
collective motions  with a relatively long wavelength along the backbone. When the ambient 
temperature increases, these collective motions cause a stepwise melting of the individual solitons,
until the backbone resembles a random chain.

Since the approach that we develop utilises only universal concepts, we conjure that it is 
similarly applicable to model the thermally induced folding and unfolding dynamics 
in a large number of different proteins. Accordingly we propose that
the folding and unfolding processes, for a large class of
proteins, should follow a {\it universal} pattern which is largely independent of the detailed
amino acid structure.  At distance scales, which are comparable to or longer than the 
C$\alpha$-C$\alpha$ virtual bond length, the Landau-Ginzburg-Wilson paradigm then seems to
become applicable and the dynamics of many proteins could
be understood in terms of universality classes.

\section{Acknowledgements}

AKS was supported in part by National Science Centre, Poland Maestro (NCN, DEC-2012/06/A/ST4/00376) and Foundation for Polish Science FNP (Mistrz7./2013) 
and by the Foundation for Polish Science (FNP START 100.2014) and  by a Swedish Institute scholarship.  AJN acknowledges support from the 
Vetenskapsr\aa det, Carl Trygger's Stiftelse f\"or vetenskaplig forskning, and Qian Ren Grant at BIT. We thank H. Scheraga for many discussions and comments. AJN also thanks G. Petsko for 
communications and suggesting the PDB structure 1ABS, P. Jennings and F. Wilczek for 
discussions, and  J. Olson and S. Kundu for communications. AJN thanks D. Melnikov and the International Institute of Physics -UFRN for hospitality during completion of this work.
Computational resources have been provided by the Informatics Center of the 
Metropolitan Academic Network (IC MAN) in Gdansk and by
the 184-processor Beowulf cluster at the Faculty of Chemistry, University of Gdansk.

\onecolumngrid
\begin{center}{
\begin{table}[h]
\begin{center}
\begin{tabular}{|l|l|l|l|l|l|p{1.6cm}|l|p{1.6cm}|}
\hline
$ $ & A & B & C & D & E & F & G & H\\ \hline
A & * & [-10,-6.5] & [-10,-7] & [-12,-9] & [-12,-8] & * & [-16,-13] & [-15,-14]\\ \hline
B & [-10,-6.5] & * & [-10,-6.5] & [-10, -7] & [-10,-7] & [-15, -12] & [-16, -13] & [-15, -12]\\ \hline
C & [-10,-7] & [-10,-6.5] & * & [-11,-6.5] & [-10,-8] & * & * & *\\ \hline
D & [-12,-9] & [-12, -6.5] & [-11,-6.5] & * & [-10,-6.5] & * & * & *\\ \hline
E & [-12,-8] & [-10,-7] & [-10,-8] & [-10,-6.5] & * & [-15,-12] & * & [-15,-13]\\ \hline
F & * & [-15,-12] & * & * & [-15,-12] & * & [-8,-5.5] & [-11,-8]\&[-15,-13]\\ \hline
G & [-16,-13] & [-16,-13] & * & * & * & [-8,-5.5] & * & [-11,-6]\\ \hline
H & [-15,-14] & [-15,-12] & * & * & [-15,-13] & [-11,-8]\&[-15,-13] & [-11,-6] & *\\
\hline
\end{tabular}
\caption{
Critical temperatures for disappearing interactions between helices. 
The symbol `` * '' indicates  that no apparent change in  contact is observed,  in the contact map
of Figures \ref{fig-13}-\ref{fig-16}.}
\label{table_2}
\end{center}
\end{table}}
\end{center}
\twocolumngrid

\end{document}